\documentclass[10pt,twocolumn,twoside]{IEEEtran}
\input epsf
\usepackage[square,numbers]{natbib}
\usepackage{xcolor,soul,framed} 
\colorlet{shadecolor}{yellow}
\usepackage{graphicx}
\usepackage{amsmath}
\newtheorem{theorem}{\textbf{Theorem}}
\newtheorem{lemma}{\textbf{Lemma}}
\newtheorem{example}{\textbf{Example}}
\newtheorem{corollary}{\textbf{{Corollary}}}
\newtheorem{remark}{\textbf{Remark}}
\newtheorem{definition}{\textbf{Definition}}

\newtheorem{assumption}{\textbf{Assumption}}
\usepackage{url}
	\newenvironment{proof}{{{\bf Proof:}}}{\hfill $\square$\par}
	\usepackage{multirow} 
	\usepackage{xcolor}
	\usepackage{multirow}
	\usepackage{setspace}
	\usepackage{url}
	\usepackage{subfigure}
	\usepackage{fancyhdr}
	\usepackage{amsmath}
	\usepackage{multirow}
	\usepackage{amssymb }
	\usepackage{color}
	\usepackage{graphics} 
	\usepackage{graphicx}
	\usepackage{algorithm} 
	\usepackage{algorithmic} 
	\usepackage{subeqnarray}
	\usepackage{cases}
	\usepackage{geometry}
	\usepackage{setspace}

	 \normalsize
\begin{document}
			
		
	\title{\LARGE Minimal Sensor Placement for Generic State and \\~ Unknown Input Observability}
			
			
			\author{Ranbo Cheng*, Yuan Zhang*, Amin MD Al, and Yuanqing Xia
				
				\thanks{This work was supported in part by the
					National Natural Science Foundation of China under Grant 62373059. *The first two authors contributed equally to this work. The authors are with School of Automation, Beijing Institute of Technology, Beijing, China.  
			Emails: {\tt\small chengranbo123@163.com, zhangyuan14@bit.edu.cn, aminmdal9723@gmail.com, xia\_yuanqing@bit.edu.cn}.}	
			 }

			\maketitle 
			{
			
			\begin{abstract}
				This paper addresses the problem of selecting the minimum number of dedicated sensors to achieve observability in the presence of unknown inputs, namely, the state and input observability, for linear time-invariant systems. We assume that the only available information is the zero-nonzero structure of system matrices, and approach this problem within a structured system model. We revisit the concept of state and input observability for structured systems, providing refined necessary and sufficient conditions for placing dedicated sensors via the Dulmage-Mendelsohn decomposition. Based on these conditions, we prove that determining the minimum number of dedicated sensors to achieve generic state and input observability is NP-hard, which contrasts sharply with the polynomial-time complexity of the corresponding problem with known inputs. We also demonstrate that this problem is hard to approximate within a factor of $(1-o(1)){\rm{log}}(n)$, where $n$ is the state dimension.  Notwithstanding, we propose nontrivial upper and lower bounds that can be computed in polynomial time, which confine the optimal value of this problem to an interval with length being the number of inputs. We further present a special case for which the exact optimal value can be determined in polynomial time. Additionally, we propose a two-stage algorithm to solve this problem approximately. Each stage of the algorithm is either optimal or suboptimal and can be completed in polynomial time.
			\end{abstract}
				\IEEEoverridecommandlockouts
				
				%
				\IEEEpeerreviewmaketitle

	
	\section{Introduction} \label{intro-sec}	
	State observability is a fundamental problem in control theory that has been extensively studied since the work of Kalman \cite{kalman1960general}. Over time, the scope of observability problem has expanded to encompass novel concepts, including state observability, state and input observability, functional observability and so on. These extensions hold significant importance in control law synthesis, fault detection and isolation, supervision and other related areas \cite{hou1992design,fernando2010functional,zhang2023observability}. Nowadays, the widespread adoption of networked control systems and cyber-physical systems, coupled with the growing concerns regarding security threats and attacks, has led to increased interest and attention in control theory research \cite{liu2011controllability,pasqualetti2015control}.
	
	In recent years, there has been a significant focus on input and output selection problems \cite{van2001review,boukhobza2009state,10098878}. One prominent issue in this domain is sensor placement to guarantee state observability in the presence of unknown inputs, namely, state and input observability (SIO). In most cases, conditions for SIO and the associated observer design heavily relied on algebraic and geometric tools, with a primary emphasis on acquiring precise knowledge of the state matrices that characterize the system's model \cite{hou1998input, tsui1996new}. However, in many modeling scenarios, obtaining accurate system parameters can be challenging, while obtaining the zero-nonzero structure is comparatively easier. Consequently, there has been a growing interest in studying system and control theory within structured system models, leveraging concepts from control theory and graph theory \cite{lin1974structural, dion2003generic}. Generic SIO (GSIO, also known as structural ISO \cite{gracy2017structural}) is the generic property corresponding to SIO studied in structured system models. GSIO implies that almost all numerical realizations of the structured system remain SIO. In recent years, sufficient and necessary conditions for GSIO have been proposed using the concepts, especially separators and vertex-disjoint edges, in directed graph (digraph) \cite{boukhobza2009state, van1999generic, boukhobza2007state}.
 Additionally, methods for sensor placement to recover GSIO based on Dulmage-Mendelsohn decomposition (DM-decomposition) have been developed, where sensors are not dedicated \cite{boukhobza2011observability}. Furthermore, many problems, like zero-dynamics attack, in cyber-physical systems and distributed systems also involve GSIO \cite{weerakkody2017robust, weerakkody2016graph, emami2020distributed}. 
	
	From a resource perspective, minimizing sensor resources to achieve system GSIO is crucial. However, many studies on the GSIO recovery problem primarily focus on utilizing undedicated sensors, that is, one sensor may measure a linear combination of an arbitrary set of state variables \cite{boukhobza2009state}, which could pose challenges in large-scale systems due to physical constraints such as geographic distance and communication capacity. Given this, the objective of this paper is to determine the minimum number of dedicated sensors (where each sensor measures only one state) required for a structured system to achieve GSIO. We call this the {\emph{minimal GSIO problem}}. As it turns out that, this problem is equivalent to determining the sparsest output matrix (a.k.a., the minimum number of connectivity links between sensors and states) to achieve GSIO. In contrast to existing work \cite{ weerakkody2017robust, weerakkody2016graph}, we focus on structured systems with a fixed topology and without the self-loop constraint. Additionally, we enforce the dedicated constraint, whereas in \cite{boukhobza2009state, boukhobza2011observability}, sensors are non-dedicated. The contributions of this paper are summarized as follows:
	\begin{itemize}
		\item[\textbullet] We rectify certain inaccuracies in a prior proposition, which is crucial for achieving GSIO using dedicated sensors \cite{boukhobza2011observability}.
		\item[\textbullet] We prove that the minimal GSIO problem is NP-hard. We further show that this problem is hard to approximate within a factor of $(1-o(1)){{\rm{log}}}(n)$, where $n$ is the state dimension. This contrasts sharply with the polynomial-time complexity of the corresponding problem with known inputs, namely, the problem of determining the minimum number of dedicated sensors to achieve structural observability (called the minimal structural observability problem \cite{pequito2015framework}) . 
		\item[\textbullet] We present an upper and a lower bound for the minimal GSIO problem when unknown inputs are dedicated, which can be computed in polynomial time. The given bounds confine the optimal value to an interval with length being the number of inputs. Central to deriving these bounds is establishing a connection between the minimal GSIO problem and the established minimal structural observability problem. We also present a special case for which the optimal value can be determined in polynomial time. 
		\item[\textbullet] We propose a two-stage heuristic algorithm for the addressed problem. Each stage is designed to be either optimal or suboptimal and can be executed in polynomial time.
	\end{itemize}
 
 
       Given that GSIO characterizes the capability to observe both states and unknown inputs, it holds significant importance in the domain of systems security, especially in the context of zero-dynamics attacks, which refer to non-zero attacks that remain stealthy for all time from the moment of input \cite{pasqualetti2013attack}. Our results emphasize the significance of sensor placement in ensuring the security of linear dynamic networks. They enable us to identify which set of states can be measured at a lower cost to effectively prevent zero-dynamics attacks.
	
	The remainder of the paper is structured as follows: Section \ref{pro-sec} presents the problem formulation. Section \ref{pre-sec} provides necessary preliminaries. In Section \ref{main-sec}, we rectify certain inaccuracies in a prior proposition and propose the corrected one. Building on this, we prove the NP-hardness of the addressed problem. Section \ref{main-sec} concludes by proposing both upper and lower bounds, along with a specific polynomial case, and introducing a two-stage algorithm. Section \ref{ex-sec} offers illustrative examples. Finally, Section \ref{con-sec} presents the conclusion.
	
	
	\section{Problem formulation}\label{pro-sec}
	In this paper, we investigate linear time-invariant systems represented by the following equations:
	
	\begin{subequations}\label{eq.system}
		\begin{equation}\label{eq.system.1}
			\dot{x}(t)=\tilde{A}x(t)+\tilde{B}u(t)
		\end{equation}
		\begin{equation}\label{eq.system.2}
			y(t)=\tilde{C}x(t)+\tilde{D}u(t)
		\end{equation}
	\end{subequations}
	where $x(t)\in \mathbb{R}^{n}$, $u(t)\in \mathbb{R}^{q}$, $y(t)\in \mathbb{R}^{m}$ are the state, unknown input and output vectors. $\tilde{A}$, $\tilde{B}$, $\tilde{C}$, and $\tilde{D}$ are numerical matrices with appropriate dimensions. The system $(\ref{eq.system})$ is given as $(\tilde{A},\tilde{B},\tilde{C},\tilde{D})$.
	
	\begin{definition} \cite{boukhobza2009state}
		The system $(\tilde{A},\tilde{B},\tilde{C},\tilde{{D}})$ is SIO if $y(t)=0$ for $t\geq 0$ implies $x(t)=0$ and $u(t)=0$ for $t\geq 0$.
	\end{definition}

	\begin{lemma}\cite{hou1999causal}\label{le.ar}
		The system $(\tilde{A},\tilde{B},\tilde{C},\tilde{D})$ is SIO if and only if its Rosenbrock matrix $R(s)$ satisfies $rank(R(s))=n+q$ for $\forall s\in \mathbb{C}$, where $R(s)=\left(\begin{matrix}\tilde{A}-s\tilde{I} & \tilde{B}\\ \tilde{C} & \tilde{D}\end{matrix}\right)$.
	\end{lemma}
     Here, $\tilde{I}$ is the identity matrix with appropriate dimension. This implies that SIO is equivalent to ensuring the system lacks invariant zeros, which is crucial for the analysis and design of controllers \cite{zhou1996robust}, and $R(s)$ maintains full column rank. 
     When considering only state observability, Lemma \ref{le.ar} reduces to the PBH test.
	
	In numerous systems, acquiring accurate state models can be challenging, while determining the zero-nonzero structures of system matrices is relatively straightforward. This paper investigates SIO within a structured system model. A structured matrix is defined as a matrix with entries that are either fixed to zero or represent free parameters that can take arbitrary values, including zero, with the flexibility for these parameters to vary independently. The set of $n_1\times n_2$ structured matrices is denoted by $\{0,*\}^{n_1\times n_2}$, where $*$ represents a free parameter. Let $M\in \{0,*\}^{n_1\times n_2}$, denote $\vert\vert M \vert\vert _0$ as the number of free parameters, which quantifies the sparsity of $M$. For two sets $\mathcal{I}\in \{1,\cdots,n_1\}$ and $\mathcal{J}\in \{1,\cdots,n_2\}$, the notation $M(\mathcal{I},\mathcal{J})$ represents the submatrix of $M$ that is composed of rows indexed by $\mathcal{I}$ and columns indexed by $\mathcal{J}$. Specially, let $M^{i}$ be the $i$th row vector of $M$, and $||M^{i}||_0$ be the number of free parameters in it. $\tilde{M}$ is a numerical matrix with the same sparsity pattern as $M$, meaning that $M_{ij}=0$ implies $\tilde{M}_{ij}=0$. $I_{n}$ represents the structured identity matrix with dimension $n$. 
	
	The structured system (\ref{eq.system}) is represented as }$(A,B,C,D)$. A numerical system $(\tilde{A},\tilde{B},\tilde{C},\tilde{D})$ is referred to as a realization of $(A,B,C,D)$. We designate a property as generic for a structured system if it holds true for almost all of its realizations. Note that ``for almost all of its realizations" implies that the structured system can adopt all parameter values within the parameter space, except for some proper algebraic variety. For the matrix $R(s)$ associated with the structured system $(A,B,C,D)$, the generic rank of $R(s)$ being $r$, denoted as $grank(R(s))=r$, implies that for almost all parameter values, $rank(R(s))=r$ for all $s\in C$. Similarly, the generic rank of a structured matrix $M$ is the rank of almost all of its realizations (which equals the maximum rank that realizations of $M$ can take \cite[Props. 2.1.12 and 2.2.25]{murota2010matrices}).
	
	\begin{definition} \cite{boukhobza2009state}
		The system $(A,B,C,D)$ is GSIO if almost all realizations of $(A,B,C,D)$ are SIO.
	\end{definition}
	
	In large-scale systems, physical constraints such as geographic distance and communication capacity often pose challenges for a single sensor to measure multiple states simultaneously. Hence, there arises a necessity to explore dedicated sensors for state measurements to solve minimal GSIO problem. As shown in Lemma \ref{lem_equ}, this problem is equivalent to finding the sparsest output matrix to achieve GSIO. We give some assumptions that will be utilized throughout this paper.

	\begin{assumption}\label{assumpall}
		\textit{1)} All inputs in this paper are considered as unknown inputs, and $B$ has full column generic rank, i.e., $grank(B)=q$. \textit{2)} The sensors placed in this paper are dedicated (except for below $\mathcal{P}^{'}$), meaning that each sensor measures only one state (i.e. $C$ is composed of some rows of $I_{n}$). \textit{3)} $D=0$ and denote the system (\ref{eq.system}) as $(A,B,C)$ (except for Corollaries \ref{corollary} and \ref{cor2}). 
	\end{assumption}

Item 3) in the above assumption indicates that sensors cannot measure unknown inputs directly. As we shall demonstrate, even when dedicated sensors are allowed to measure unknown inputs, the considered problem ${\cal P}$ below remains NP-hard  (see Corollary \ref{corollary}). Therefore, for conciseness, we consider the system $(A,B,C)$ with $D=0$. In  Corollaries \ref{corollary} and \ref{cor2}, we shall discuss extending our results to the case $D\ne 0$. The full column generic rank of $B$ ensures that the following addressed problem is feasible when $D=0$, which is common in the literature \cite{darouach1994full,weerakkody2016graph}.   

	\textit{\textbf{Problem} ($\mathcal{P}$):} Minimal GSIO problem of $(A,B)$:
		\begin{equation}\label{Pro}
			\begin{split}
				&\min_{C\in \{0,*\}^{n\times n}} \vert\vert C\vert\vert_0\\
				s.t.&(A,B,C) \text{ is GSIO.}\\
				&\vert \vert C^i\vert \vert_0\leq 1\text{ for }1\leq i\leq n.
			\end{split}
		\end{equation}
  
	Given that $B$ has full column generic rank, $C=I_n$ is always a feasible solution to the above problem. Therefore, problem ${\mathcal P}$ is well-defined. A related problem to ${\mathcal P}$ is finding the sparsest output matrix to achieve GSIO, formulated as follows. 
 
	$\mathcal{P}^{'}$: Sparsest sensor placement to achieve GSIO of $(A,B)$:
		\begin{equation}\label{Pro2}
			\begin{split}
				&\min_{C\in \{0,*\}^{n\times n}} \vert\vert C\vert\vert_0\\
				s.t.&(A,B,C) \text{ is GSIO.}
			\end{split}
		\end{equation}
	    \begin{lemma}\label{lem_equ}
		    Problems $\mathcal{P}$ and $\mathcal{P}^{'}$ have the same optimal value.
	    \end{lemma}
	    
	    \begin{proof}
	    	On the one hand, the optimal sensor placement for $\mathcal{P}$ is a feasible solution for $\mathcal{P}^{'}$. On the other hand, suppose $\mathcal{Y}^{'}$ is the optimal sensor placement for $\mathcal{P}^{'}$, and $C^{'}$ is its corresponding output matrix. Let $\mathcal{X}^{'}$ be the set of states measured by sensors of $\mathcal{Y}^{'}$. We construct a sensor placement $\mathcal{Y}$ with output matrix being $C$, where each $x_i\in \mathcal{X}^{'}$ is measured by a dedicated sensor belonging to $\mathcal{Y}$. Denote $R^{'}(s)$ and $R(s)$ as the Rosenbrock matrices for $(A,B,C^{'})$ and $(A,B,C)$, respectively. Observe that $grank(R^{'}(s))=n+q$ implies $grank(R(s))=n+q$ for $\forall s\in C$. Thus, $(A,B,C)$ achieves GSIO, and $C$ is a feasible solution for $\mathcal{P}$. Therefore, $\mathcal{P}$ and $\mathcal{P}^{'}$ have the same optimal value.
	    \end{proof}
		
	\section{Preliminaries}\label{pre-sec}
	In this section we introduce some preliminaries in graph theory. Digraph $\mathcal{G}(\mathcal{V},\mathcal{E})$ is utilized to represent the structured system $(A,B,C)$. Here, $\mathcal{V}=X\cup U\cup Y$, where $X=\{x_1,\cdots,x_n\}$, $U=\{u_1,\cdots,u_q\}$ and $Y=\{y_1,\cdots, y_m\}$ represent the sets of state vertices, input vertices and output vertices, separately. The edge set $\mathcal{E}=\mathcal{E}_{xx}\cup \mathcal{E}_{ux}\cup \mathcal{E}_{xy}$, where $\mathcal{E}_{xx}=\{(x_j,x_i)|A(i,j)\neq 0\}$, $\mathcal{E}_{ux}=\{(u_j,x_i)|B(i,j)\neq 0\}$ and $\mathcal{E}_{xy}=\{(x_j,y_i)|C(i,j)\neq 0\}$. Here, $M(i,j)$ represents the $(i,j)$ entry of the matrix $M$, and $(x_j,x_i)$ denotes a directed edge from $x_j$ to $x_i$. Moreover, $(x_j,x_i)$ is referred to as the outgoing edge (resp. incoming edge) of $x_j$ (resp. $x_i$). Two edges $(x_j,x_i)$ and $(x_{j'},x_{i'})$ are said to be vertex-disjoint if $x_j\neq x_{j'}$ and $x_i\neq x_{i'}$. Given two subsets $\mathcal{V}_1$ and $\mathcal{V}_2$ of $\mathcal{V}$. $\theta (\mathcal{V}_1,\mathcal{V}_2)$ denotes the maximum number of vertex-disjoint edges from $\mathcal{V}_1$ to $\mathcal{V}_2$. A path from $v_i$ to $v_{j}$ is a sequence of edges $\{(v_i,v_{i+1}),(v_{i+1},v_{i+2}),\cdots ,(v_{j-1},v_j)\}$ without repeated vertices. A digraph is said to be strongly connected if, for any two vertices of it, there exist paths from each vertex to the other one. A subgraph of $\mathcal{G}$ is considered a strongly connected component (SCC) if it is strongly connected and no additional edges or vertices can be included without breaking the property. A path is considered a $\mathcal{V}_1-\mathcal{V}_2$ path if its starting vertex belongs to $\mathcal{V}_1$ and ending vertex belongs to $\mathcal{V}_2$. We denote $\rho (\mathcal{V}_1,\mathcal{V}_2)$ as the maximum number of mutually disjoint $\mathcal{V}_1-\mathcal{V}_2$ paths. A set of $\rho (\mathcal{V}_1,\mathcal{V}_2)$ mutually disjoint $\mathcal{V}_1-\mathcal{V}_2$ paths is referred to as maximum $\mathcal{V}_1-\mathcal{V}_2$ linking. We define $\mathcal{V}_{ess}(\mathcal{V}_1,\mathcal{V}_2)$ as the set of vertices $v\in \mathcal{V}$ covered by any maximum $\mathcal{V}_1-\mathcal{V}_2$ linking. A state vertex $x_i$ is said to be $Y$-reached if there exists a path from $x_i$ to some $y_i\in Y$ in $\mathcal{G}(\mathcal{V},\mathcal{E})$.
	
\begin{figure}[t]
			\centering
			\includegraphics[width=8.5cm]{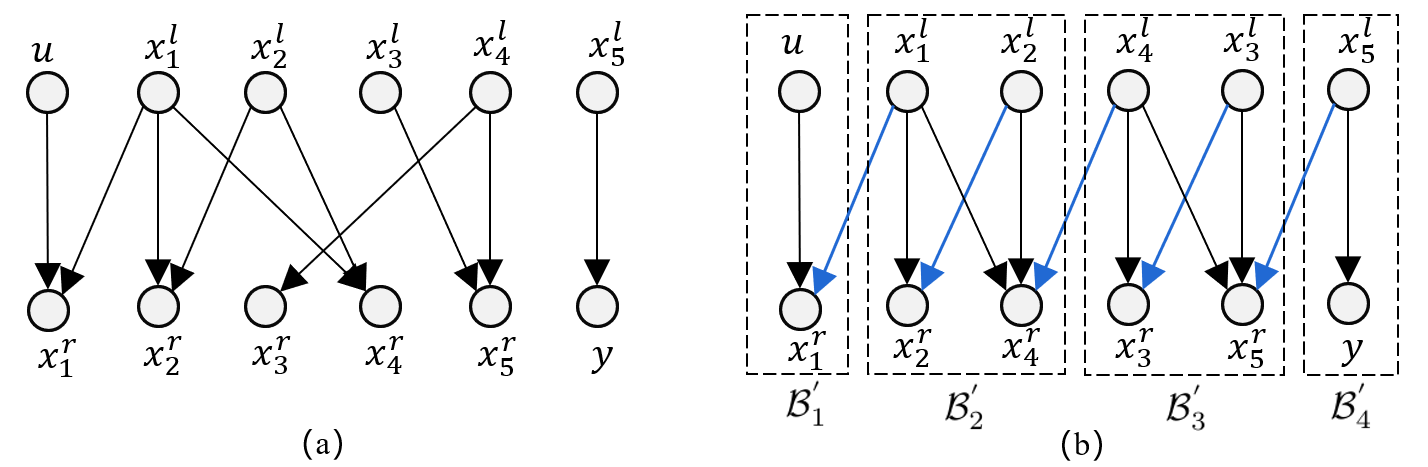}
			\caption{(a) The bipartite digraph of $(A,B,C)$ in Example \ref{ex.bd}. (b) $\mathcal{D}(\mathcal{B}^{'}(A,B,C))$ of Example \ref{ex.bd} with 4 strongly connected components, where blue lines represent \textit{s-edges}.}
			\label{fig.bd}
		\end{figure}

	A bipartite digraph, denoted as $\mathcal{B}(A,B,C)=(\mathcal{V}^{l},\mathcal{V}^{r},\mathcal{E}^{'})$, is associated with the structured system $(A,B,C)$, where $\mathcal{V}^{l}$ and $\mathcal{V}^{r}$ are two disjoint vertex sets, and $\mathcal{E}^{'}$ is the edge set from $\mathcal{V}^{l}$ to $\mathcal{V}^{r}$. More precisely, $\mathcal{V}^{l}=X^{l}\cup U$ and $\mathcal{V}^{r}=X^{r}\cup Y$, where $X^{l}=\{x_1^{l},\dots ,x_n^{l}\}$ and $X^{r}=\{x_1^r,\dots ,x_n^{r}\}$ are two copies of state vertices. The edge set $\mathcal{E}^{'}=\mathcal{E}^{'}_{xx}\cup \mathcal{E}^{'}_{ux}\cup \mathcal{E}^{'}_{xy}$, where $\mathcal{E}^{'}_{xx}=\{(x_j^l,x_i^r)|A(i,j)\neq 0\}$, $\mathcal{E}^{'}_{ux}=\{(u_j,x_i^r)|B(i,j)\neq 0\}$, $\mathcal{E}^{'}_{xy}=\{(x_j^l,y_i)|C(i,j)\neq 0\}$. In a bipartite digraph, a matching is defined as a set of mutually vertex-disjoint edges. The size of a matching is the number of edges it contains. The matching with the largest size among all matchings in a bipartite digraph is referred to as the maximum matching. Within a matching, a vertex is considered right-matched (resp. left-matched) if it serves as an end vertex belonging to $\mathcal{V}^{r}$ (resp. $\mathcal{V}^{l}$) of an edge within the matching; otherwise, it is right-unmatched (resp. left-unmatched). A right-perfect (resp. left-perfect) matching in a bipartite digraph $\mathcal{B}=(\mathcal{V}^l,\mathcal{V}^r,\mathcal{E}^{'})$ is a matching that covers every vertex in $\mathcal{V}^r$ (resp. $\mathcal{V}^l$). The cardinality of a set $\mathcal{V}$ is denoted as $card(\mathcal{V})$. Hereafter, we present an example to illustrate the bipartite digraph representation.
	\begin{example}\label{ex.bd}
		For a structured system $(A,B,C)$ with
		$$A=\left[\begin{matrix}
			*&0&0&0&0\\
			*&*&0&0&0\\
			0&0&0&*&0\\
			*&*&0&0&0\\
                0&0&*&*&0
		\end{matrix}\right]
		B=\left[\begin{matrix}
			*\\
			0\\
			0\\
			0\\
                0
		\end{matrix}\right]
		C=\left[\begin{matrix}
			0\\
			0\\
			0\\
			0\\
                *
		\end{matrix}\right]^{T}$$
	\end{example}
		The bipartite digraph representing it is depicted in Fig. \ref{fig.bd} (a).
	
	The following part provides a brief introduction to DM-decomposition \cite{murota2010matrices}. For a maximum matching $\mathcal{M}$ in $\mathcal{B}=(\mathcal{V}^l,\mathcal{V}^r,\mathcal{E}^{'})$, let $\partial^{l}\mathcal{M}$ (resp. $\partial^{r}\mathcal{M}$) denote the set of vertices in $\mathcal{V}^l$ (resp. $\mathcal{V}^r$) covered by the edges of $\mathcal{M}$. We define $\mathcal{S}^l=\mathcal{V}^l\backslash\partial^{l} \mathcal{M}$ and $\mathcal{S}^r=\mathcal{V}^r\backslash\partial ^{r} \mathcal{M}$. An auxiliary bipartite graph $\tilde{\mathcal{B}}=(\mathcal{V}^l,\mathcal{V}^r,\tilde{\mathcal{E}})$ is defined, where $\tilde{\mathcal{E}}=\{(u,v)|(u,v)\in \mathcal{E}^{'}$ or $(v,u)\in \mathcal{M}\}$. 
    The steps of DM-decomposition follow subsequently. For a maximum matching $\mathcal{M}$, let $\mathcal{V}_{0}=\{v\in \mathcal{V}^{l}\cup \mathcal{V}^{r}|$ $\exists$ a path from $u$ to $v$ in $\tilde{\mathcal{B}}$ for some $u \in \mathcal{S}^{l} \}$, and $\mathcal{V}_{\infty}=\{v\in \mathcal{V}^{l}\cup \mathcal{V}^{r}|$ $\exists$ a path from $v$ to $u$ in $\tilde{\mathcal{B}}$ for some $u\in \mathcal{S}^{r}\}$. Denote $\tilde{\mathcal{B}}^{'}$ as the subgraph of $\tilde{\mathcal{B}}$ obtained by deleting the vertices $\mathcal{V}_0\cup \mathcal{V}_{\infty}$ and edges incident to them. Let $\tilde{\mathcal{B}}_i (i=1,\dots ,k)$ denote the SCCs of $\tilde{\mathcal{B}}^{'}$, and $\mathcal{V}_i$ be the vertex sets corresponding to $\tilde{\mathcal{B}}_i$. Denote $\mathcal{B}_i$ as the subgraphs of $\mathcal{B}$ induced on $\mathcal{V}_i(i=0,1,2,\cdots ,\infty)$. Define a partial order $``\prec"$ on $\{\mathcal{B}_i|i=1,\cdots,k\}$, where $\mathcal{B}_{i_1}\prec \mathcal{B}_{i_2}$ if there exists a path on $\tilde{\mathcal{B}}$ from $v_{i}\in \mathcal{V}_{i_2}$ to $v_{j}\in \mathcal{V}_{i_1}$. Also, define $\mathcal{B}_0\prec \mathcal{B}_{i}\prec \mathcal{B}_{\infty}$ for any $i\in \{1,\dots,k\}$. The order of the subscripts of $\mathcal{B}_i(i=1,\cdots,k)$ follows this partial order. 
    Thus, the decomposed graph of $\mathcal{B}$ by DM-decomposition is denoted as $\mathcal{D}(\mathcal{B})=(\mathcal{B}_0,\mathcal{B}_1,\cdots,\mathcal{B}_k,\mathcal{B}_{\infty})$.
    Each vertices set $\mathcal{V}_i$ can be divided into two parts, $\mathcal{V}_i^l=\mathcal{V}_i\cap \mathcal{V}^l$ and $\mathcal{V}_i^r=\mathcal{V}_i\cap \mathcal{V}^r$. It is important to note that the obtained subgraphs $\mathcal{B}_i$ remain the same regardless of the choice of the maximum matching $\mathcal{M}$ \cite{murota2010matrices}.
	
	For the structured system $(A,B,C)$, we denote a bipartite digraph $\mathcal{B}^{'}(A,B,C)=(\mathcal{V}^l,\mathcal{V}^r,\mathcal{E}\cup \mathcal{E}_s)$ corresponding to $R(s)$, where $\mathcal{E}_s=\{(x^l_i,x^r_i)|i=1,\cdots n\}$. $\mathcal{D}(\mathcal{B}^{'}(A,B,C))=(\mathcal{B}^{'}_{0},\mathcal{B}^{'}_{1},\cdots ,\mathcal{B}^{'}_{k},\mathcal{B}^{'}_{\infty})$ is the DM-decomposition of it.  Note that no parallel edges are introduced even if $\mathcal{E}\cap \mathcal{E}_s\neq \emptyset$. The edges in $\mathcal{E}_s$ are referred to as \textit{s-edges}. Fig. \ref{fig.bd} (b) illustrates the DM-decomposition $\mathcal{D}(\mathcal{B}^{'}(A,B,C))$ of Example \ref{ex.bd} for clarity.
	
	\section{Main results}\label{main-sec}
	\subsection{Conditions of Sensor Placement for GSIO} 
	In this subsection, we review some sufficient and necessary conditions of GSIO utilizing graph theory and provide refined propositions for $\mathcal{P}$.
	
	\begin{lemma}\cite[Proposition 4]{boukhobza2011observability}\label{le.cond}
	 The system $(A,B,C)$ is GSIO if and only if the following two conditions hold simultaneously
		\begin{itemize} 
			\item[\textit{1)}] $\mathcal{D}(\mathcal{B}(A,B,C))$ does not contain $\mathcal{B}_0$ (i.e. $\mathcal{V}_0=\emptyset$).
			\item[\textit{2)}] In $\mathcal{D}(\mathcal{B}^{'}(A,B,C))$, $\mathcal{B}^{'}_{i}$ does not contain \textit{s-edges} for $\forall i\in\{1,\cdots ,k\}$.
		\end{itemize}
	\end{lemma}

	Lemma \ref{le.cond} suggests that the sensor placement for $\mathcal{P}$ can be divided into two steps corresponding to the two conditions. In \textit{step 1}, states are selected for measurement (i.e. designing $C_1$) to achieve \textit{1)}. Based on $(A,B,C_1)$, \textit{step 2} selects other states for measurement (i.e. designing $C_2$) to achieve \textit{2)}. A method for placing minimal sensors to satisfy \textit{1)} is proposed in the following Lemma.
	
	\begin{lemma}\cite[Proposition 5]{boukhobza2011observability}\label{le.1)}
		Consider the system $(A,B)$ and the corresponding $\mathcal{D}(\mathcal{B}(A,B))$. To satisfy \textit{1)} of Lemma \ref{le.cond}, the minimum number of sensors is $\gamma=card(\mathcal{V}_0^{l})-card(\mathcal{V}_0^{r})$. These $\gamma$ sensors, denoted as $C_1$, must measure states such that the maximum matching $\mathcal{M}$ of $\mathcal{B}(A,B,C_1)$ is left-perfect.
	\end{lemma}
	
	Then, we revisit \textit{Proposition 7} from \cite{boukhobza2011observability}, which proposes that achieving \textit{2)} necessitates additional sensors to measure at least one state in each $\mathcal{B}_{i}^{'}$ that contains \textit{s-edges}, where $i\in\{1,\cdots,k\}$. However, for Example \ref{ex.bd} which has satisfied \textit{1)} of Lemma \ref{le.cond}, placing only one sensor to measure one state of $\{x_1,x_2\}$ can transform $\mathcal{B}_2^{'}$, $\mathcal{B}_{3}^{'}$ and $\mathcal{B}_{4}^{'}$ to $\mathcal{B}_{\infty}^{'}$, and there are no \textit{s-edges} in $\mathcal{B}_{1}^{'}$. Thus, we do not need to measure at least one state in each $\mathcal{B}_{i}^{'}$ that contains \textit{s-edges}. 
 
 Inspired by the analysis of \cite{boukhobza2011observability}, since any two vertices in the same strongly connected component remain strongly connected after adding a new sensor anywhere, the only way to place sensors to achieve \textit{2)} of Lemma \ref{le.cond} is to transform all $\mathcal{B}_{i}^{'}, i\in \{1\cdots,k\}$, containing \textit{s-edges} to $\mathcal{B}_{\infty}^{'}$. We then provide the following revised theorem. 
        Let $[C_1;C_2]$ be the matrix formed by stacking $C_1$ and $C_2$. Define $\mathcal{R}_{i}=\{x_k|x_k^l\in \mathcal{V}_{i}^{l} \cup \mathcal{V}_{j}^{l}$, where $\mathcal{B}_{j}^{'}\prec \mathcal{B}_{i}^{'}\}$ corresponding to $\mathcal{B}_{i}^{'}$ in $\mathcal{D}(\mathcal{B}^{'}(A,B,C_1))$.
	\begin{theorem}\label{th.2)}
		Consider the system $(A,B,C_1)$ and the corresponding $\mathcal{D}(\mathcal{B}^{'}(A,B,C_1))$ which satisfies \textit{1)} of Lemma \ref{le.cond}. In order to achieve \textit{2)}, additional sensors must measure at least one state in $\mathcal{R}_i$ if $\mathcal{B}_i^{'}$ contains \textit{s-edges}.
	\end{theorem}
	\begin{proof}
		Firstly, we demonstrate the sufficiency of the theorem. After \textit{step 1}, there is no $\mathcal{B}_0^{'}$ in $\mathcal{D}(\mathcal{B}^{'}(A,B,C_1))$. For any $\mathcal{B}_i^{'}(i\in \{1,\cdots,k\})$ that contains \textit{s-edges}, place a sensor $y_{i}$ to measure any vertex in $\mathcal{R}_{i}$ corresponding to $\mathcal{B}_i^{'}$. Since choosing different maximum matching does not impact DM-decomposition, we still choose the maximum matching $\mathcal{M}$ used in \textit{step 1}. Then, $y_{i}\in \mathcal{S}^r$. According to DM-decomposition, $\mathcal{B}_{i}^{'}$ is transformed into $\mathcal{B}_{\infty}^{'}$ because there exists a path from any $v_i\in \mathcal{B}_i^{'}$ to $y_{i}$ in the auxiliary bipartite graph of $\mathcal{B}^{'}(A,B,[C_1;C_2])$. Therefore, after placing sensors for all $\mathcal{B}_i^{'}$ containing \textit{s-edges}, every $\mathcal{B}_{i}^{'}(i=1,\cdots,k)$ in $\mathcal{D}(\mathcal{B}^{'}(A,B,[C_1;C_2]))$ does not contain \textit{s-edges}.
		
		Next we demonstrate the necessity of the theorem. For any $\mathcal{B}_{i}^{'}$ that contains \textit{s-edges} in $\mathcal{D}(\mathcal{B}^{'}(A,B,C_1))$, suppose that additional sensors measure vertices in $\mathcal{B}_{j}^{'}$ where $\mathcal{B}_{j}^{'} \nprec \mathcal{B}_{i}^{'}$ (i.e., there is no path from vertex in $\mathcal{B}_{i}^{'}$ to $\mathcal{B}_{j}^{'}$). We choose the same maximum matching $\mathcal{M}$ used in \textit{step 1}. At this time, the new $\mathcal{S}^{r'}$ is composed of original $\mathcal{S}^{r}$, $y_i$ and the vertices in $\mathcal{B}_{l}^{'}$, where $\mathcal{B}_{j}^{'}\prec \mathcal{B}_{l}^{'}$. For any $v_i\in \mathcal{B}_{i}^{'}$, there does not exist a path from $v_i$ to any vertex in $\mathcal{S}^{r'}$. $\mathcal{B}_{i}^{'}$ still contains \textit{s-edges}. Therefore, it is necessary to measure at least one state in $\mathcal{R}_i$.
	\end{proof}

	
	\subsection{Complexity Analysis}
	Theorem \ref{th.2)} is useful for determining whether the system is GSIO. However, obtaining the minimum sensor placement is challenging due to potential overlaps among different $\mathcal{R}_i$. This subsection is dedicated to proving that $\mathcal{P}$ is NP-hard. First we introduce the so-called extended set cover problem.
	\begin{definition}\label{def.extended}
		Consider $q$ nonempty subsets $\mathcal{S}_i(i=1,\cdots ,q)$ of $\mathcal{S}=\{1,\cdots ,p\}$ such that every element of $\mathcal{S}$ belongs to at least one subset $\mathcal{S}_i(i\in \{1,\cdots ,q\})$. Consider other $p$ subsets $\mathcal{S}_{q+i}=\{i\}(i=1,\cdots, p)$ of $\mathcal{S}$. The extended set cover problem is to find the minimal number of $\mathcal{S}_i(i\in \{1,\cdots, q+p\})$ such that their union covers $\mathcal{S}$.
	\end{definition}
	
	\begin{lemma}\label{le.extended}
		The extended set cover problem is NP-hard.
	\end{lemma}
	\begin{proof}
		This lemma is proven by employing the set cover problem, a well-known NP-hard problem \cite{hochbaum1982approximation}. Denoting an arbitrary set cover problem as $\mathcal{Q}_1$. In $\mathcal{Q}_1$, we consider $q$ nonempty subsets $\mathcal{S}_i$ of $\mathcal{S}=\{1,\cdots,p\}$ such that every element of $\mathcal{S}$ appears in at least one set $\mathcal{S}_i(i\in\{1,\cdots,q\})$. The objective of $\mathcal{Q}_1$ is to determine the minimal number of $\mathcal{S}_i$ whose union is $\mathcal{S}$. Based on the formulation of $\mathcal{Q}_1$, we construct an extended set cover problem $\mathcal{Q}_2$. In $\mathcal{Q}_2$, we consider $q$ nonempty subsets $\mathcal{W}_i$ of $\mathcal{W}$, where $\mathcal{W}_i(i=1,\cdots,q)$ and $\mathcal{W}$ are the same to $\mathcal{S}_i(i=1,\cdots,q)$ and $\mathcal{S}$. Additionally, we introduce other $p$ subsets $\mathcal{W}_{q+i}=\{i\}(i=1,\cdots,p)$. The objective of $\mathcal{Q}_2$ is to find the minimal number of $\mathcal{W}_i$ whose union is $\mathcal{W}$. $\mathcal{Q}_2$ can be efficiently constructed from $\mathcal{Q}_1$ in polynomial time. 
		
		We demonstrate that $\mathcal{Q}_1$ has a solution with cardinality $k$ if and only if $\mathcal{Q}_2$ has a solution with cardinality $k$.
		
		Suppose that $\mathcal{Q}_2$ has a solution $\mathcal{L}_2=\{\mathcal{W}_i|\cup \mathcal{W}_i=\mathcal{W}\}$ with cardinality $k$. Define $\mathcal{L}_2^1=\{\mathcal{W}_i|\mathcal{W}_i\in \mathcal{L}_2 \text{ and } 1\leq i\leq q\}$, $\mathcal{L}_2^2=\{\mathcal{W}_i|\mathcal{W}_i\in \mathcal{L}_2 \text{ and } q+1\leq i\leq q+p\}$ and $\mathcal{L}_2^3=\{\mathcal{W}_i|\mathcal{W}_i\notin \mathcal{L}_2 \text{ and }1\leq i\leq q\}$. According to the construction of $\mathcal{Q}_2$, each element of $\mathcal{W}$ appears in at least one set $\mathcal{W}_i(i\in\{1,\cdots,q\})$. Thus, for any $\mathcal{W}_j\in \mathcal{L}_2^2$, there exists $\mathcal{W}_i\in \mathcal{L}_2^3$ such that $\mathcal{W}_j \subseteq \mathcal{W}_i$. Moreover, $\mathcal{L}_2$ is the minimal solution, meaning that there is no any other $\mathcal{W}_h\in \mathcal{L}_2^2$ where $h\neq j$ such that $\mathcal{W}_h\subseteq \mathcal{W}_i$. Therefore, $\mathcal{W}_j\in \mathcal{L}_2$ can be replaced by $\mathcal{W}_i$, resulting in a new solution $\mathcal{L}_2^{'}$ of $\mathcal{Q}_2$ with cardinality $k$. Note that if more than one element of $\mathcal{L}_2^3$ contains the same element of $\mathcal{L}_2^2$, any of them can be chosen. As a consequence, define a set $\mathcal{L}_1=\{\mathcal{S}_i|\mathcal{S}_i \text{ has the same subscript with } \mathcal{W}_i\in \mathcal{L}_2^{'}\}$ with cardinality $k$, and it is a solution of the set cover problem $\mathcal{Q}_1$. 
		
	 On the other hand, let $\mathcal{L}_1=\{\mathcal{S}_i|\cup \mathcal{S}_i=\mathcal{S}\}$ with cardinality $k$ be a solution of $\mathcal{Q}_1$. According to the construction of $\mathcal{Q}_2$, define a set $\mathcal{L}_2=\{\mathcal{W}_i|\mathcal{W}_i \text{ has the same subscript with } \mathcal{S}_i\in \mathcal{L}_1\}$ with cardinality $k$. It is observed that $\mathcal{L}_2$ is a solution to $\mathcal{Q}_2$ due to $\mathcal{S}=\mathcal{W}$ and $\mathcal{S}_i=\mathcal{W}_i$ for $i=1,\cdots, q$.
		
		Since the set cover problem is NP-hard, it follows that the extended set cover problem is NP-hard, too. 
	\end{proof}
	
Observe that Lemma \ref{le.cond} imposes two conditions, where the viability of \textit{2)} is heavily contingent on the sensor placement established in \textit{1)}. To mitigate the interdependence between \textit{1)} and \textit{2)} when designing the minimum sensor placement, we establish that a subclass of $\mathcal{P}$, denoted as $\mathcal{P}_{1}$, is NP-hard. $\mathcal{P}_{1}$ represents a subclass of $\mathcal{P}$ wherein the minimum sensor placement of \textit{1)} is uniquely determined based on Lemma \ref{le.1)}. This implies that the redundant sensors identified in \textit{step 1} can be treated as sensors for \textit{step 2}. Consequently, the sensor placement required to achieve \textit{2)} in $\mathcal{P}_{1}$ remains unaffected by the variations in \textit{step 1}, ensuring a more streamlined and independent design process.
	
	\begin{theorem}\label{th.NP}
		Problem $\mathcal{P}$ is NP-hard. Provided that $P\neq NP$, $\mathcal{P}$ cannot be approximated within a factor of $(1-o(1)){\rm{log}}$$(n)$, where $n$ denotes the state dimension.
	\end{theorem}
	\begin{proof}
		The proof is based on a reduction from the extended set cover problem. Consider an extended set cover problem on $q+p$ subsets $\{{\cal S}_i|_{i=1}^{q+p}\}$ of ${\cal S}=\{1,...,p\}$ given in Definition \ref{def.extended}, denoted as $\mathcal{Q}_1$, we construct $\mathcal{P}_{1}$, a subclass of $\mathcal{P}$ as follows. Initially, we denote that ``$s$" as a free parameter in a structured matrix, referring to the entry corresponding to an \textit{s-edge} in bipartite digraph. We then construct a structured matrix $M(s)\in \{0,*,s\}^{(2p+3q)\times (2p+3q)}$ based on $\mathcal{Q}_1$, where $M(s)$ can be permuted into the Rosenbrock matrix to determine a structured system of $\mathcal{P}_1$. Moreover, the diagonal matrix of $M(s)$ corresponds to each $\mathcal{B}_{i}^{'}$ of the DM-decomposed bipartite digraph of the constructed system. The construction of $M(s)$ is carried out through the following steps (see (\ref{example.M}) for illustration):
		\begin{itemize}
			\item [1.] The diagonal entries of $M(s)$ are $*$.
			\item [2.] $(i,i+1)$ and $(i+1,i)$ entries of $M(s)$ are $s$, where $i=2q+1,2q+3,\cdots,2q+2p-1$.
			\item [3.] $(i,i+q)$ and $(i+q,i+2q+2p)$ entries of $M(s)$ are $s$, where $i=1,2,\cdots,q$.
			\item [4.] $(i+q,2j+2q-1)$ entries of $M(s)$ are $*$ when $S_{q+j}\subseteq S_{i}$ in $\mathcal{Q}_1$, where $1\leq i\leq q$ and $1\leq j\leq p$.
			\item [5.] Other entries of $M(s)$ are 0.
		\end{itemize}
		
	    After permuting $M(s)$, we obtain $R(s)$ whose first $2p+2q$ diagonal entries are $s$. Let $\mathcal{I}=\{1,2,\cdots,2p+2q\}$ and $\mathcal{J}=\{2p+2q+1,\cdots,2p+3q\}$, define structured matrices $A=R(s)(\mathcal{I},\mathcal{I})-sI_{2p+2q}$, $B=R(s)(\mathcal{I},\mathcal{J})$, and $C_1=R(s)(\mathcal{\mathcal{J},\mathcal{I}})$. 
        Thus, we obtain $\mathcal{P}_1$ with $(A,B)$. In more detail, $\mathcal{P}_1$ possesses a unique minimum dedicated sensor placement $C_1$ for $(A,B)$ to achieve \textit{1)} of Lemma \ref{le.cond}. This is because there must be no zero columns in the Rosenbrock matrix, i.e., $R(s)$, of $(A,B,C_1)$. Consequently, $\vert \vert C_1\vert\vert_0$ is a constant. $\mathcal{P}_1$ can be constructed from $\mathcal{Q}_1$ in polynomial time utilizing the outlined steps. 
		
		Notably, the DM-decomposition $\mathcal{D}(\mathcal{B}^{'}(A,B,C_1))$ with $(A,B,C_1)$ above is the same as the bipartite graph of $M(s)$, where $\mathcal{V}^{l}$ is composed of the column vertices of $M(s)$ and $\mathcal{V}^{r}$ is composed of the row vertices of $M(s)$. Observe that the subgraph $\mathcal{B}^{'}_{i}(i= 1,2,\cdots,p+3q)$ of $\mathcal{D}(\mathcal{B}^{'}(A,B,C_1))$ corresponds to the $i$th (from top left to bottom right) block diagonal submatrix of $M(s)$ with the form of either $\left( \begin{matrix}*\end{matrix}\right)$ or $\left(\begin{matrix}*&s\\s&*\end{matrix}\right)$ (Note that there is no $\mathcal{B}^{'}_0$ and $\mathcal{B}^{'}_{\infty}$ in $\mathcal{D}(\mathcal{B}^{'}(A,B,C_1))$). To facilitate further discussion, define $\mathcal{W}=\{\mathcal{B}^{'}_i|i=2q+1,2q+2,\cdots,2q+p\}$, representing the set of subgraphs containing the \textit{s-edge} in $\mathcal{D}(\mathcal{B}^{'}(A,B,C_1))$. $\mathcal{W}$ corresponds to $\mathcal{S}$ in $\mathcal{Q}_1$ with $\mathcal{B}_{i+2q}^{'}\in \mathcal{W}$ corresponding to $i\in \mathcal{S}$ for $i=1,\cdots, p$. Additionally, define $\mathcal{L}=\{\mathcal{B}^{'}_{i}|i=q+1,q+2,\cdots,2q+p\}$, representing the set of subgraphs where sensors can be placed. $\mathcal{L}$ corresponds to the set of subsets $\mathcal{S}_i$ in $\mathcal{Q}_1$ with $\mathcal{B}_{i+q}^{'}\in \mathcal{L}$ corresponding to $\mathcal{S}_i$ for $i=1,\cdots,q+p$. Notably, placing only one sensor in each $\mathcal{B}^{'}_i\in \mathcal{L}$ is sufficient, any excess is redundant.
		
		We claim that $\mathcal{Q}_1$ has a solution with cardinality $k$ if and only if $\mathcal{P}_{1}$ has a minimum sensor placement with size $k+\vert\vert C_1\vert\vert_0$.
		
		Suppose that $\mathcal{P}_{1}$ can be solved with $k+\vert\vert C_1\vert\vert_0$ sensors, where $k$ sensors represent the minimum required for achieving $(A,B,C_1)$ to satisfy \textit{2)}. These $k$ sensors must be placed in $k$ different $\mathcal{B}^{'}_{i+q}\in \mathcal{L},i \in \{1,\cdots,q+p\}$, otherwise, there is a placement with a smaller size. The $k$ subgraphs $\mathcal{B}^{'}_{i+q}$ form a solution set denoted as $\mathcal{H}_1$. Let $\mathcal{H}_2$, with cardinality $k$, be a set of subsets $\mathcal{S}_{i}$ in $\mathcal{Q}_1$,  where each element $\mathcal{S}_{i}$ corresponds to $\mathcal{B}^{'}_{i+q}\in \mathcal{H}_1$ (having the same $i$). According to Theorem \ref{th.2)}, for each $\mathcal{B}^{'}_{i+q}\in \mathcal{W}$, there is a $\mathcal{B}^{'}_{j}\in \mathcal{H}_1$ such that $\mathcal{B}^{'}_{j}\prec \mathcal{B}^{'}_{i+q}$ or $\mathcal{B}^{'}_{i+q}=\mathcal{B}^{'}_j$. Therefore, according to \textit{step 4} in the construction above, for every element in $\mathcal{S}$, there is a subset $\mathcal{S}_{j}\in \mathcal{H}_2$ containing it. Consequently, $\mathcal{H}_{2}$ is a solution of $\mathcal{Q}_1$.
		
		Conversely, suppose that $\mathcal{Q}_1$ has a solution $\mathcal{H}_2=\{\mathcal{S}_i|\cup \mathcal{S}_i=\mathcal{S}\}$ with cardinality $k$. Let $\mathcal{H}_1$, with cardinality $k$, be a set of subgraphs of $\mathcal{D}(\mathcal{B}^{'}(A,B,C_1))$ in $\mathcal{P}_1$, whose elements are $\mathcal{B}^{'}_{i+q}$ corresponding to $\mathcal{S}_i\in \mathcal{H}_2$ (having the same $i$), $i\in\{1,\cdots,q+p\}$. As shown in \textit{step 4} in the above construction, for any $\mathcal{B}^{'}_{j}\in \mathcal{W}$, there exists at least one $\mathcal{B}^{'}_{i+q}\in \mathcal{H}_1$ such that $\mathcal{B}^{'}_{i+q}=\mathcal{B}^{'}_{j}$ or $\mathcal{B}^{'}_{i+q}\prec \mathcal{B}^{'}_{j}$, due to any element of $\mathcal{S}$ belonging to at least one $\mathcal{S}_{i}\in \mathcal{H}_2$. This demonstrates that we can place a sensor to measure one state in every $\mathcal{B}^{'}_{i+q}\in \mathcal{H}_1$ to achieve \textit{2)} of Lemma \ref{le.cond}. Then we can place $k+\vert\vert C_1\vert\vert_0$ sensors to solve $\mathcal{P}_{1}$, where $\vert\vert C_1\vert\vert_0$ sensors are used to satisfy \textit{1)}, and $k$ sensors for \textit{2)}.
		
        As a result of the above, $\mathcal{P}_{1}$ is NP-hard. Therefore, $\mathcal{P}$ is NP-hard. Moreover, the proof of Lemma \ref{le.extended} implies that the size of the optimal solution to the extended set cover problem is the same as that of the set cover problem. Additionally, the size of the optimal solution to problem $\mathcal{P}_1$ is $||C_1||_0$ more than the size of the solution to the extended set cover problem. Since $(1-o(1)){\rm{log}}$$(n)$ is a threshold below which the set cover problem cannot be efficiently approximated \cite{feige1998threshold}, $\mathcal{P}_1$ is inapproximable to a factor of $(1-o(1)){\rm{log}}$$(n)$ due to the linear relations between the solution to $\mathcal{P}_1$ and the set cover problem. Consequently, problem $\mathcal{P}$ cannot be approximated within a factor of $(1-o(1)){\rm{log}}$$(n)$.
	\end{proof}
	
	An example is provided to enhance the clarity of this proof.
	
	\begin{example}\label{example_NP}
		Consider an extended set cover problem with the set $S=\{1,2,3\}$. The subsets of $S$ are $S_1=\{1,2\}$, $S_2=\{2,3\}$, $S_3=\{1,3\}$, $S_4=\{1\}$, $S_5=\{2\}$, $S_6=\{3\}$. We then construct the structured matrix $M(s)$ as follows. 
	{\small{ \begin{equation} \label{example.M}
			\setlength{\arraycolsep}{4pt}
         M(s)=\left[\begin{array}{ccccccccccccccc}
				*&0&0&s&0&0&0&0&0&0&0&0&0&0&0\\
				0&*&0&0&s&0&0&0&0&0&0&0&0&0&0\\
				0&0&*&0&0&s&0&0&0&0&0&0&0&0&0\\
				0&0&0&*&0&0&*&0&*&0&0&0&s&0&0\\
				0&0&0&0&*&0&0&0&*&0&*&0&0&s&0\\
				0&0&0&0&0&*&*&0&0&0&*&0&0&0&s\\
				0&0&0&0&0&0&*&s&0&0&0&0&0&0&0\\
				0&0&0&0&0&0&s&*&0&0&0&0&0&0&0\\
				0&0&0&0&0&0&0&0&*&s&0&0&0&0&0\\
				0&0&0&0&0&0&0&0&s&*&0&0&0&0&0\\
				0&0&0&0&0&0&0&0&0&0&*&s&0&0&0\\
				0&0&0&0&0&0&0&0&0&0&s&*&0&0&0\\
				0&0&0&0&0&0&0&0&0&0&0&0&*&0&0\\
				0&0&0&0&0&0&0&0&0&0&0&0&0&*&0\\
				0&0&0&0&0&0&0&0&0&0&0&0&0&0&*\\
			\end{array} \right]
		\end{equation}}}
  After permuting $M(s)$, we obtain $R(s)$, 

{\small \begin{equation}\label{rosen}
\setlength{\arraycolsep}{4pt}
R(s)=\left[\begin{array}{cccccccccccc|ccc}
                s&0&0&0&0&0&0&0&0&0&0&0&*&0&0\\
				0&s&0&0&0&0&0&0&0&0&0&0&0&*&0\\
				0&0&s&0&0&0&0&0&0&0&0&0&0&0&*\\
				*&0&0&s&0&0&0&*&0&*&0&0&0&0&0\\
				0&*&0&0&s&0&0&0&0&*&0&*&0&0&0\\
				0&0&*&0&0&s&0&*&0&0&0&*&0&0&0\\
				0&0&0&0&0&0&s&*&0&0&0&0&0&0&0\\
				0&0&0&0&0&0&*&s&0&0&0&0&0&0&0\\
				0&0&0&0&0&0&0&0&s&*&0&0&0&0&0\\
				0&0&0&0&0&0&0&0&*&s&0&0&0&0&0\\
                0&0&0&0&0&0&0&0&0&0&s&*&0&0&0\\
				0&0&0&0&0&0&0&0&0&0&*&s&0&0&0\\\hline
                0&0&0&*&0&0&0&0&0&0&0&0&0&0&0\\
				0&0&0&0&*&0&0&0&0&0&0&0&0&0&0\\
				0&0&0&0&0&*&0&0&0&0&0&0&0&0&0\\
\end{array}\right]
\end{equation}}
  
			

\begin{figure}[t]
	 	\centering
	 	\includegraphics[width=9cm]{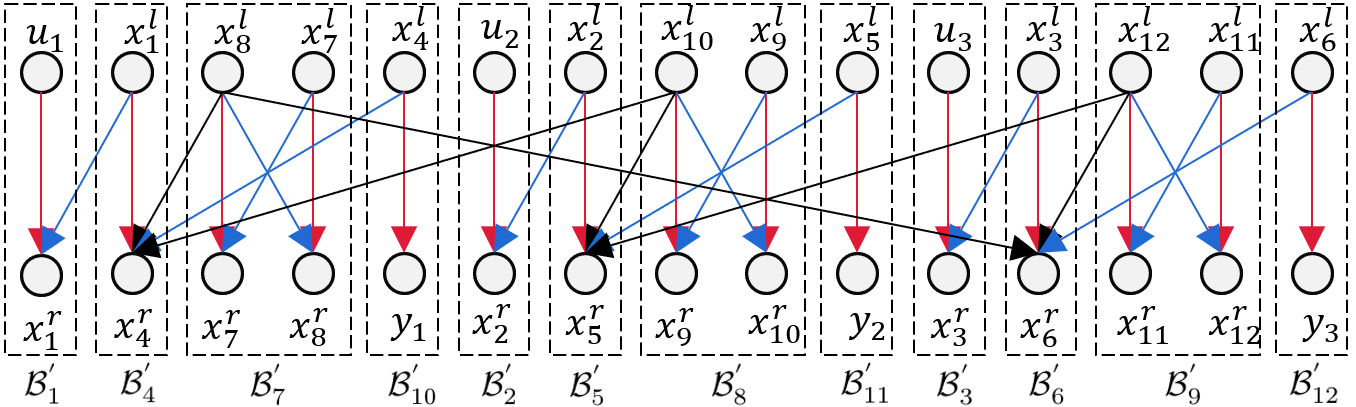}
	 	\caption{ The bipartite digraph of $\mathcal{D}(\mathcal{B}^{'}(A,B,C_{1}))$ in Example \ref{example_NP}, where red lines represent maximum matching and blue lines represent \textit{s-edges}.}
	 	\label{fig.7}
	 \end{figure}
  
  Thus, $\mathcal{P}$ is obtained with the system whose Rosenbrock matrix is (\ref{rosen}), featuring the unique minimum sensor placement for \textit{1)} in it. $M(s)$ has 12 block diagonal submatrices corresponding to $\mathcal{B}^{'}_1$ to $\mathcal{B}^{'}_{12}$ in $\mathcal{D}(\mathcal{B}^{'}(A,B,C_{1}))$, arranged from the top-left to bottom-right, as depicted in Fig. \ref{fig.7}. The set $\mathcal{S}$ in the extended set cover problem corresponds to $\mathcal{W}=\{\mathcal{B}^{'}_7,\mathcal{B}^{'}_8,\mathcal{B}^{'}_9\}$ with elements containing \textit{s-edges}. The subsets $\mathcal{S}_1$ to $\mathcal{S}_6$ correspond to $\mathcal{B}^{'}_4$ to $\mathcal{B}^{'}_9$ respectively. According to Theorem \ref{th.2)}, placing a sensor to measure one state of $\mathcal{B}_4^{'}$ (resp. $\mathcal{B}_5^{'}$ or $\mathcal{B}_6^{'}$) transforms $\mathcal{B}_{7}^{'}$ and $\mathcal{B}_8^{'}$ (resp. $\mathcal{B}_8^{'}$ and $\mathcal{B}_9^{'}$, or $\mathcal{B}_9^{'}$ and $\mathcal{B}_7^{'}$) to $\mathcal{B}_{\infty}^{'}$. Similarly, placing a sensor to measure one state of $\mathcal{B}_i^{'}$ transforms $\mathcal{B}_{i}^{'}$ to $\mathcal{B}_{\infty}^{'}$ for $i=7,8,9$. Thus, placing minimum sensors in $\{\mathcal{B}^{'}_4,\mathcal{B}^{'}_5,\cdots,\mathcal{B}^{'}_9\}$ to achieve GSIO can determine the minimal number of subsets $\mathcal{S}_i$ such that their union is $\mathcal{S}$.
	\end{example}
	
		\begin{remark} \label{np-discussion}
			The NP-hardness of $\mathcal{P}$ and $\mathcal{P}^{'}$ are in sharp contrast with the similar problem for known inputs, i.e., the minimal structural observability problem. It has been found that the latter problem, as well as its various variants, is polynomially solvable \cite{pequito2015framework, 10098878}. It follows from Theorem \ref{th.NP} that determining minimal dedicated sensors to prevent zero-dynamics attacks is also NP-hard.
		\end{remark}
		
		

The following corollary implies that even when dedicated sensors can be deployed to directly measure unknown inputs, problem $\mathcal{P}$ is still NP-hard. More precisely, consider the following problem $\mathcal{P^{''}}$ without \textit{3)} of Assumption \ref{assumpall}, and the structured system is denoted as $(A,B,C,D)$.

  $\mathcal{P}^{''}$: Minimal GSIO problem allowing direct measurement of unknown inputs: 
		\begin{equation}\label{Pro_D}
			\begin{split}
				&\min_{[C,D]\in \{0,*\}^{(n+q)\times (n+q)}} \vert\vert C\vert\vert_0+\vert\vert D\vert\vert_0\\
				s.t.&(A,B,C,D) \text{ is GSIO.}\\
				&\vert \vert [C^i\quad D^i]\vert \vert_0\leq 1\text{ for }1\leq i\leq n+q.
			\end{split}
		\end{equation}

        \begin{corollary}\label{corollary}
Problem $\mathcal{P^{''}}$ is NP-hard. Provided that $P\neq NP$, $\mathcal{P}^{''}$ is inapporximable to a factor $(1-o(1)){\rm{log}}$$(n)$, where $n$ denotes the state dimension.
        \end{corollary}

        \begin{proof}
            The corollary can be directly derived from the proof of Theorem \ref{th.NP}. Utilizing the same construction outlined in Theorem \ref{th.NP}, it follows that the same $C_{1}$ and $D_1=0$ are also the unique minimum sensors ensuring \textit{1)} of Lemma \ref{le.cond}, as there must be no zero columns in $R(s)$ of $(A,B,C_1,D_1)$. Subsequently, since $D$ can be designed, We adjust $\mathcal{L}$ to $\mathcal{L}^{'}=\{\mathcal{B}_{i}^{'}|i=1,\cdots,2q+p\}$, while keeping $\mathcal{W}$ unchanged. Placing a sensor in $\mathcal{B}_{i}^{'}$ indicates the measurement of the input $u_{i}$ for $i\in\{1,\cdots,p\}$. According to Theorem \ref{th.2)}, observations from $\mathcal{D}(\mathcal{B}^{'}(A,B,C_1,D_1))$ indicate that placing a sensor in either $\mathcal{B}_{i}^{'}$ or $\mathcal{B}_{q+i}^{'}$ serves identical functions for $i\in\{1,\dots,q\}$, since $\mathcal{B}_{i}^{'}$ has only one incoming edge, which is from $\mathcal{B}_{q+i}^{'}$. Hence, placing a sensor in $\mathcal{B}_{i}^{'}$ can be substituted by placing a sensor in $\mathcal{B}_{q+i}^{'}$ for $i\in \{1,\cdots,q\}$ without increasing the number of sensors. Consequently, according to Theorem \ref{th.NP}, $\mathcal{P}^{''}$ is NP-hard and cannot be approximated within a factor of $(1-o(1)){\rm{log}}$$(n)$, where $n$ denotes the state dimension.
        \end{proof}
	
	\subsection{Upper and Lower Bounds}\label{subsection.c}
	Despite the NP-hardness of problem $\mathcal{P}$, this subsection provides upper and lower bounds for $\mathcal{P}$ that can be computed in polynomial time, relating $\mathcal{P}$ to the minimal structural observability problem. 
	
	\begin{assumption}
		All inputs considered in this subsection are dedicated, implying that each input exclusively drives a single state, and conversely, each state is driven by at most one input (i.e., $B$ is composed of certain rows of $I$).
	\end{assumption}

\begin{figure}[t]
	 	\centering
	 	\includegraphics[width=8cm]{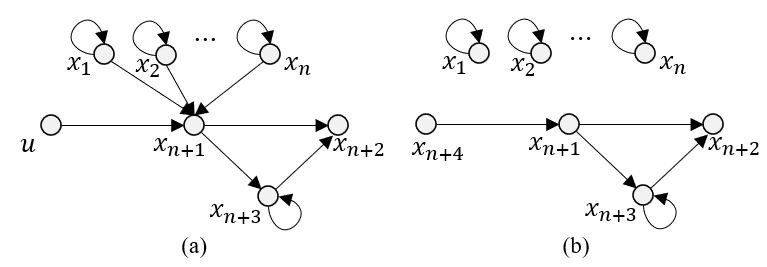}
	 	\caption{(a) The digraph of $(A,B)$ in Example \ref{ex.bounds}. (b) The digraph of the auxiliary system corresponding to $\hat{A}$ in Example \ref{ex.bounds}.}
	 	\label{fig.bounds}
	 \end{figure}

 Consider a system $(A,B)$ with $q$ dedicated inputs. Let $\mathcal{G}_1$ be its digraph representation.  Define a vertex set $\mathcal{I}\doteq \{x_i|x_i$ has an incoming edge from $u_j$ in $\mathcal{G}_1$, $j\in\{1,\cdots,q\}\}$, with $card(\mathcal{I})=q$. Additionally, define an associated auxiliary digraph $\hat{\mathcal{G}}_1$, which is a subgraph of $\mathcal{G}_1$ obtained by removing the edges from any state vertex $x_i$ to $x_j$ where $x_j\in \mathcal{I}$. By considering input vertices of $\hat{\mathcal{G}}_1$ as state vertices, we obtain a digraph, whose associated structured matrix is given by $\hat{A}$. We call the system associated with $\hat{A}$ the  auxiliary system (to simplify notations, we may use $\hat A$ to denote the auxiliary system). See Fig. \ref{fig.bounds} for the illustration of $\hat A$ corresponding to a system $(A,B)$.  
 

The following two lemmas give necessary and sufficient conditions for structural observability and GSIO,  respectively, where $B$ and $C$ need not be dedicated. 
	\begin{lemma}\cite[Theorem]{lin1974structural}\label{le.obsv}
		The system $(A,B,C)$ is structural observable if and only if its digraph $\mathcal{G}(\mathcal{V},\mathcal{E})$ satisfies the following conditions:
		\begin{itemize}
			\item[\textit{1)}] $\theta (X,X\cup Y)=n$.
			\item[\textit{2)}] Every $x_i\in \mathcal{V}$ is $Y$-reached.
		\end{itemize}
	\end{lemma}
	
	\begin{lemma}\cite[Proposition 4]{boukhobza2009state}\label{le.3cond}
		The system $(A,B,C)$ is GSIO if and only if its digraph $\mathcal{G}(\mathcal{V},\mathcal{E})$ satisfies the following conditions:
		\begin{itemize}
			\item[\textit{1)}] $\theta (X\cup U,X\cup Y)=n+q$.
			\item[\textit{2)}] Every $x_i\in \mathcal{V}$ is $Y$-reached.
			\item[\textit{3)}] $\Delta_0\subseteq \mathcal{V}_{ess}(U,Y)$ where $\Delta_0=\{x_i|\rho (U\cup \{x_i\},Y)=\rho (U,Y)\}$.
		\end{itemize}
	\end{lemma}
	
	Define $H(\hat{A})$ as the optimal value of the minimal structural observability problem associated with $\hat A$, i.e., $H(\hat{A})$ is the minimum number of dedicated sensors required for $\hat{A}$ to achieve structural observability. $H(\hat{A})$ can be computed in $O(n^{2.5})$ time (c.f. Remark \ref{np-discussion}).
	 
	 \begin{theorem}\label{th.bounds}
	 	For a structured system $(A,B)$ with $q$ dedicated inputs, the optimal value of $\mathcal{P}$ is in the integer interval $[H(\hat{A}),H(\hat{A})+q]$.
	 \end{theorem}
	 \begin{proof}
	 	Firstly, we prove that $H(\hat{A})$ serves as a lower bound. Assume that $C$ is a sensor placement for $(A,B)$ such that $(A,B,C)$ is GSIO. Denote $\mathcal{G}_1^{'}$ as the digraph of $(A,B,C)$ and $\hat{\mathcal{G}}_1^{'}$ as the digraph of $(\hat{A},C)$. We need to show that $(A,B,C)$ being GSIO implies that $(\hat{A},C)$ achieves structural observability. Since the inputs are dedicated to states, each edge $(u_j,x_i)$ must be contained in the $n+q$ vertex-disjoint edges from $U\cup X$ to $X\cup Y$ in $\mathcal{G}_{1}^{'}$. Therefore, edges from any state $x_j$ to $x_i\in \mathcal{I}$ do not belong to the $n+q$ vertex-disjoint edges. Consequently, $\hat{\mathcal{G}}_1^{'}$ satisfies \textit{1)} of Lemma \ref{le.obsv}. Additionally, assume that $\hat{\mathcal{G}}_1^{'}$ does not satisfy \textit{2)} of Lemma \ref{le.obsv}. This implies that there exists some $x_i$ that is not $Y$-reached, or each path from $x_i$ to $y_i$ contains $x_j\in \mathcal{I}$ in $\mathcal{G}_{1}^{'}$. In the first case, $\mathcal{G}_{1}^{'}$ does not satisfy \textit{2)} of Lemma \ref{le.3cond}. In the second case, it indicates that $x_j\in \mathcal{V}_{ess}(\{x_i\}, Y)$. Let $u_j$ be the input vertex that links to $x_j$, we have $x_{j}\in \mathcal{V}_{ess}(\{u_j\},Y)$ since the input $u_j$ is dedicated. Then, we have $\rho(\{x_i,u_j\},Y)=1$, which implies $x_i\in \Delta_0$. Moreover, suppose $x_i\in \mathcal{V}_{ess}(U,Y)$. This implies that there always exists a path from an input $u_k$ different from $u_j$ (since $u_j$ links to $x_j$) to some $y_i$ (denoted as $(u_k, \cdots, x_i,x_j, \cdots, y_i$)) such that this path is contained in the maximum $U-Y$ linking. However, for this maximum linking, we can change the path $(u_k, \cdots, x_i,x_j, \cdots, y_i$) to $(u_j,x_j,\cdots y_i)$, where the orders of the edges after $x_j$ are the same as the former one, such that there is a $U-Y$ linking with the same size that does not contain $x_i$, which contradicts the assumption. Thus, $x_i\notin \mathcal{V}_{ess}(U,Y)$. Therefore, $\mathcal{G}_{1}^{'}$ dose not satisfy \textit{3)} of Lemma \ref{le.3cond}, indicating that $(A,B,C)$ is not GSIO, which contradicts our initial assumption. Therefore, every $x_i$ is $Y$-reached in $\hat{\mathcal{G}}_1^{'}$, indicating that $(\hat{A},C)$ is structurally observable. Consequently, $H(\hat{A})$ serves as a lower bound for $\mathcal{P}$. 
	 	
	 	Next, we prove that $H(\hat{A})+q$ serves as an upper bound. Assume that $C$ is a sensor placement for $\hat{A}$, such that $(\hat{A},C)$ achieves structural observability, with $\vert\vert C\vert\vert_0=H(\hat{A})$. In the digraph $\hat{\mathcal{G}}_{1}^{'}$ associated with $(\hat{A},C)$, denote $\mathcal{O}\subseteq \mathcal{I}$, where $x_i\in \mathcal{O}$ if $x_{i}$ is measured by a sensor and has outgoing edges to other state vertices. Additionally, define $E_i^{o}$ as the set of vertices that serve as an end vertex (different from $x_i$) of the outgoing edge of $x_i$. For each $x_i\in \mathcal{O}$, let $x_j\in E_{i}^{o}$ be the end vertex of the edge $(x_l,x_j)$, where $(x_l,x_j)$ is contained in the $n+q$ vertex-disjoint edges that satisfy condition 1) of Lemma \ref{le.obsv} for $\hat{\mathcal{G}}_{1}^{'}$. Now, we can position the sensor to measure $x_l$ instead of $x_i$ to obtain $C^{'}$ with $\vert \vert C^{'}\vert\vert_0=\vert\vert C\vert \vert _0=H(\hat{A})$, such that $(\hat{A},C^{'})$ achieves structural observability. Note that if all such $x_l$ corrsponding to $x_i$ are measured, the sensor remains measuring this $x_i$. Moreover, we denote $C^{*}=[C^{'};C^{''}]$, where $C^{''}$ represents $p(p\leq q)$ sensors measuring $x_i\in \mathcal{I}$, which are not measured by $C^{'}$, with $\vert \vert C^{*}\vert\vert_0=H(\hat A)+p\leq H(\hat A)+q$. Then, the digraph of $(A,B,C^{*})$ is denoted as $\mathcal{G}_{2}^{'}$. Observe that $\mathcal{G}_{2}^{'}$ satisfies \textit{1)} and \textit{2)} of Lemma \ref{le.3cond}. Since the inputs are dedicated, all vertices of $\mathcal{I}$ belong to $\mathcal{V}_{ess}(U,Y)$. Additionally, each $x_i\in \mathcal{I}$ is measured by dedicated sensors, and all $x_j\notin \mathcal{I}$ are $Y$-reached in $\mathcal{G}_2^{'}$, implying that $\rho(U\cup \{x_j\},Y)=\rho(U,Y)+1$. Thus, there is no state vertex belonging to $\Delta_0$ except for $\mathcal{I}$, and all vertices of $\mathcal{I}$ belong to $\mathcal{V}_{ess}(U,Y)$. Therefore, $\mathcal{G}_{2}^{'}$ satisfies \textit{3)} of Lemma \ref{le.3cond}. $(A,B,C^{*})$ achieves GSIO. Therefore, $H(\hat{A})+q$ serves as the upper bound for $\mathcal{P}$.
	 \end{proof}
	 
	 	The upper and lower bounds are non-trivial. In the worst case, the lower bound $H(\hat{A})$ can be of size $(n+q)$. Despite this, the length of the interval between the given bounds is $q$, which could be orders of magnitude smaller than $n$. Thus, the upper and lower bounds significantly reduce the search space for finding the optimal solution to the NP-hard ${\cal P}$. Additionally, if $H(\hat{A})+q>n$, the upper bound is adjusted to $n$. The following example shows the importance of these bounds.
	 \begin{example}\label{ex.bounds}
	 	The upper and lower bounds of $(A,B)$ depicted in (a) of Fig. \ref{fig.bounds} are $[n+1,n+2]$, where $n+1$ represents the optimal value of the minimal structural observability problem for $\hat{A}$, and the number of input is 1. Actually, we need $n+2$ sensors to measure $x_1,\cdots,x_n,x_{n+2},x_{n+3}$ to achieve GSIO in this example, meeting the given bounds. This example illustrates that the number of sensors needed for GSIO is much more than that for structural observability, which needs only one sensor to measure $x_{n+2}$. Therefore, the provided bounds greatly reduce the searching space for GSIO.
	 \end{example}

      Inspired by Theorem \ref{th.bounds}, if sensors can be deployed to directly measure inputs, an analogous upper bound and lower bound is given by the following corollary, where both bounds are smaller than those in Theorem \ref{th.bounds}, and the length is also the number of inputs.
      \begin{corollary}\label{cor2}
          For a structured system $(A,B)$ with $q$ dedicated inputs, the optimal value of $\mathcal{P}^{''}$ is in the integer interval $[H(A),H(A)+q]$.
      \end{corollary}
      \begin{proof}
          The lower bound $H(A)$ can be easily obtained due to conditions \textit{1)} and \textit{2)} of Lemma \ref{le.3cond}. For the upper bound, we construct a sensor placement where $q$ sensors measure the $q$ inputs respectively, and the other $H(A)$ sensors form any minimal senor placement for the minimal structural observability problem of $A$. For this $H(A)+q$ sensor placement, \textit{1)} and \textit{2)} of Lemma \ref{le.3cond} hold. Moreover, since $\mathcal{V}_{ess}(U,Y)=U$, we have $\Delta_0=\emptyset$. Thus, \textit{3)} of Lemma \ref{le.3cond} is satisfied. Consequently, $H(A)+q$ is the upper bound.
      \end{proof}

\begin{remark}
     The difference between Theorem \ref{th.bounds} and Corollary \ref{cor2} arises from the fact that prohibiting sensors from directly measuring unknown inputs requires that the state nodes belonging to $\mathcal{I}$ must match some input nodes, while this is not required when sensors can directly measure inputs. Based on Theorem \ref{th.bounds}, the crucial problem of determining sensor bounds lies in finding $H(\hat{A})$, which can be solved using various methods proposed in \cite{pequito2015framework, 10098878}. Thus, the bounds can be computed in time $O(n^3)$ with $n$ being the dimension of system.
\end{remark}
  
	 Despite $\mathcal{P}$ being NP-hard, Theorem \ref{th.bounds} provides an interval to search for the optimal solution of $\mathcal{P}$. Particularly, when the number of unknown inputs is small, this theorem can offer a relatively accurate sensor placement regardless of the  size of system. However, when inputs are not dedicated, the minimum number of sensors does not meet the lower bound. 

\subsection{A Polynomially Solvable Case}

 \begin{figure}[t]
	 	\centering
	 	\includegraphics[width=9cm]{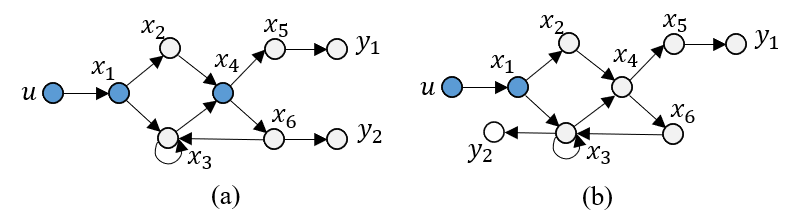}
	 	\caption{(a) and (b) represent the same structured system with two different minimum sensor placements to achieve structural observability. Blue vertices belong to $\mathcal{V}_{ess}(U,Y)$. (b) is GSIO but (a) is not.}
	 	\label{fig.6}
	 \end{figure}

A question naturally arises following Theorem \ref{th.bounds}, that is, whether ${\mathcal{P}}$ can be solved efficiently when there is only one dedicated input (i.e., $q=1$)? The difficulty stems from \textit{3)} of Lemma \ref{le.3cond}. More precisely, different $Y$ causes varying $\mathcal{V}_{ess}(U,Y)$, as depicted in Fig. \ref{fig.6}, resulting in the same number of sensors leading to diverse GSIO properties of the system. This subsection discusses how we can efficiently solve $\mathcal{P}$ when every state vertex of the system digraph has a self-loop.

Consider a structured system denoted as $(A^{'}, e_{i})$ with one dedicated input $u$, where $A^{'}_{ii}=*, i=1,\cdots, n$ (implying that every state vertex has a self-loop), and $e_{i}$ represents the input matrix corresponding to the unique dedicated input $u$. The digraph of $(A^{'},e_{i})$ is denoted as $\mathcal{G}_{1}$. Let $\hat{A}^{'}$ be the auxiliary system of $(A^{'},e_{i})$ defined in Subsection \ref{subsection.c}, and $\hat{\mathcal{G}}_1$ be the digraph of $(\hat{A}^{'},C)$. Let $\hat{Y}_{min}$ be the minimum sensor placement, with the output matrix being $\hat{C}_{min}$, such that $(\hat{A}^{'},\hat{C}_{min})$ is structurally observable (noting that there may be different $\hat{Y}_{min}$ and $\hat{C}_{min}$). Denote $\mathcal{G}_{1}^{'}$ and $\hat{\mathcal{G}}_{1}^{'}$ as the digraph of $(A^{'},e_i,\hat{C}_{min})$ and $(\hat{A}^{'},\hat{C}_{min})$, respectively. Denote $L(\hat{A}^{'})$ as the number of input reachable sink-SCCs of $\hat{A}^{'}$, where an SCC is considered a sink-SCC if there are no outgoing edges from its vertices to other SCCs. Input-reachable means that there exists a path from an input $u$ to states of this SCC. From the definition we have $L(\hat{A}^{'})\leq H(\hat{A}^{'})$. Then, a lemma is proposed which is crucial for the sensor placement of this special case.

\begin{lemma}\label{lem.selfloop}
    Given the system $(A^{'},e_{i},\hat{C}_{min})$, the vertices set $\mathcal{V}_{ess}(\{u\},\hat{Y}_{min})$ in $\hat{\mathcal{G}}_{1}^{'}$ remains constant across different minimum sensor placements $\hat{Y}_{min}$ and $\hat{C}_{min}$ when $L(\hat{A}^{'})>1$.
\end{lemma}

\begin{proof}
    Given that every state vertex has a self-loop in $\hat{\mathcal{G}}_{1}$ and there is only one input vertex $u$, only condition \textit{2)} of Lemma \ref{le.obsv} is required to achieve structural observability for $\hat{A}^{'}$. \textit{2)} of Lemma \ref{le.obsv} is equivalent to ensuring that each sink-SCC has at least one state vertex to be measured by sensors. Since $L(\hat{A}^{'})>1$, $\hat{\mathcal{G}}_{1}$ has $k$ sink-SCCs, where $k>1$. Let $X_{i},i=1,\cdots,k$, denote the sets of state vertices for each sink-SCC. Then, for any minimum sensor placement $\hat{Y}_{min}^{1}$ and $\hat{C}_{min}^{1}$, $(\hat{A}^{'},\hat{C}_{min}^{1})$ being structurally observable implies $X_i \cap \mathcal{V}_{ess}(\{u\},\hat{Y}_{min}^{1})=\emptyset, \forall i\in\{1,\cdots,k\}$ in $\hat{\mathcal{G}}_{1}^{'}$. This is because $u$ has paths to $L(\hat{A}^{'})$ different sink-SCCs. Suppose $\hat{Y}_{min}^{2}$ and $\hat{C}_{min}^{2}$ represent another minimum sensor placement such that $(\hat{A}^{'},\hat{C}_{min}^{2})$ is structurally observable, we have $\mathcal{V}_{ess}(\{u\},\hat{Y}_{min}^{1})=\mathcal{V}_{ess}(\{u\},\hat{Y}_{min}^{2})$ in $\hat{\mathcal{G}}_1^{'}$. This equality arises because if we consider $\hat{\mathcal{G}}_{2}^{'}$ as the subgraph of $\hat{\mathcal{G}}_{1}^{'}$ obtained by removing any $x_{i}\in \mathcal{V}_{ess}(\{u\},\hat{Y}_{min}^{1})$, there exists no path from $u$ to any $y\in \hat{Y}_{min}^{1}$. Thus, there is also no path from $u$ to any state vertex $x_{i}\in \bigcup _{i=1}^{k}X_{i}$ in $\hat{\mathcal{G}}_{2}^{'}$. Therefore, there is also no path from $u$ to any $y\in \hat{Y}_{min}^{2}$ in $\hat{\mathcal{G}}_{2}^{'}$, as $\hat{Y}_{min}^{2}$ is the minimum sensor placement and solely measures states belonging to the sink-SCCs. Consequently, the $\mathcal{V}_{ess}(\{u\},\hat{Y}_{min}^{i})$ remains constant for any minimum sensor placement $\hat{Y}_{min}^{i}$.
\end{proof}

Lemma $\ref{lem.selfloop}$ ensures that $\mathcal{V}_{ess}(\{u\},\hat{Y}_{min})$ remains unchanged with different $\hat{Y}_{min}$, simplifying the minimum sensor design for \textit{3)} of Lemma \ref{le.3cond}. Based on Lemma \ref{lem.selfloop},  $\mathcal{V}_{ess}(\{u\},\hat{Y}_{min})$ is used in the subsequent discussion regardless of the specific $\hat{Y}_{min}$ is. Denote the state vertex driven by $u$ as $x^{*}$. The minimum sensor placement can be obtained by Algorithm \ref{alg.self}.

\begin{algorithm}[b]
		\caption{: Polynomial-time algorithm for $\mathcal{P}$ of $(A^{'},e_i)$}\label{alg.self}
		\begin{algorithmic}[1]
			\REQUIRE The structured system $(A^{'},e_i)$ and its digraph $\mathcal{G}_{1}$ ($A'_{ii}=*$, $i=1,...,n$)
			\ENSURE Minimum sensor placement $Y_{min}$
			\STATE Construct the auxiliary system $\hat{A}^{'}$ and its digraph $\hat{\mathcal{G}}_1$.\label{step1}
                \STATE Find the minimum sensor placement $\hat{Y}_{min}$ (with output matrix being $\hat{C}_{min}$), such that $(\hat{A}^{'},\hat{C}_{min})$ is structurally observable.\label{step2}
			\STATE Find the $\mathcal{V}_{ess}(\{u\},\hat{Y}_{min})$ in $\hat{\mathcal{G}}_1^{'}$, where $\hat{\mathcal{G}}_{1}^{'}$ is the digraph of $(\hat{A}^{'}, \hat{C}_{min})$.\label{step3}
			\FOR{$i=1$ to $n$} \label{step4}
			\IF{$x_{i}\notin \mathcal{V}_{ess}(\{u\},\hat{Y}_{min})$ and $\rho (\{u,x_{i}\},\hat{Y}_{min})=\rho(\{u\},\hat{Y}_{min})$ in $\hat{\mathcal{G}}_{1}^{'}$}\label{step5}
			\RETURN $Y_{min}=\hat{Y}_{min}\cup \{y^{*}\}$, where $y^{*}$ is a sensor measuring $x^{*}$.
			\ENDIF
			\ENDFOR
			\RETURN $Y_{min}=\hat{Y}_{min}$
		\end{algorithmic}
	\end{algorithm}

\begin{theorem}\label{th.selfloop}
    Algorithm \ref{alg.self} is correct and can be completed in time $O(n^{3.5})$.
\end{theorem}

\begin{proof}
    Firstly, we prove the correctness of Algorithm \ref{alg.self}. The $\hat{Y}_{min}$ is necessarily required to satisfy \textit{1)} and \textit{2)} of Lemma \ref{le.3cond}. Then, if $\hat{Y}_{min}$ fails to satisfy \textit{3)} of Lemma \ref{le.3cond}, according to the proof of Theorem \ref{th.bounds}, an additional sensor $y^{*}$ is required. Therefore, Algorithm \ref{alg.self} is correct. 

    Next, we analyze the complexity of Algorithm \ref{alg.self}. Step \ref{step1} constructs the auxiliary system $\hat{A}^{'}$. It can be completed in time $O(M)=O(n^2)$, where $M$ is the maximum possible number of edges in $\mathcal{G}_{1}$. In worst case $M=n^2+n$. Step \ref{step2} determines $\hat{Y}_{min}$. Since every state has a self-loop in $\hat{\mathcal{G}}_{1}$, it can be transformed into strongly connected components decomposition, achievable through an algorithm with a complexity order $O(Nlog(N)=O(nlog(n)))$ \cite{fleischer2000identifying}, where $N$ is the number of vertices in $\hat{\mathcal{G}}_1$, and in worst case, $N=n+1$. In step \ref{step3}, the computation of $\mathcal{V}_{ess}(\{u\},\hat{Y}_{min})$ can be solved by solving a Maximum-Flow problem with a complexity of $O({N^{'}}^{0.5}M^{'})=O(n^{2.5})$ \cite{ahuja1995network}, where $N^{'}$ and $M^{'}$ are the number of vertices and edges in $\hat{\mathcal{G}}_{1}^{'}$, respectively. In worst case, $N^{'}=n+1+card(\hat{Y}_{min})$ and $M^{'}=n^2+n+n\times card(\hat{Y}_{min})+card(\hat{Y}_{min})$. Similarly, Step \ref{step4} requires $n$ computations of Maximum-Flow problem, with a complexity of $n\times O({N^{'}}^{0.5}M^{'})=O(n^{3.5})$. Consequently, the global complexity of Algorithm \ref{alg.self} is $O(n^{3.5})$.
\end{proof}

\begin{remark}
    In the case where the system features only one dedicated input and each state vertex has a self-loop, Theorem \ref{th.selfloop} can efficiently solve $\mathcal{P}$ without the need to search in large space, whose size is $\tbinom{n}{H(A^{'})}+\tbinom{n}{H(A^{'}+1)}$.
\end{remark}

\begin{remark}
    When $L(\hat{A}^{'})=1$, Lemma \ref{lem.selfloop} is not applicable. In this case, a simple traversal approach described as follows can be employed to determine $Y_{min}$, where $Y_{min}=Y_{min}^1\cup Y_{min}^{2}$ with $card(Y_{min}^{2})\leq 2$. Here, $Y_{min}^{1}$ measures each sink-SCC in $\hat{\mathcal{G}}_{1}$ that cannot be reached by $u$. This can be obtained in time $O(nlog(n))$. Then, we search every $x_{i}$ and determine $Y_{min}^{2}$ such that $(A^{'},e_i,C_{min})$ satisfy conditions in Lemma \ref{le.3cond}. The proof of Theorem \ref{th.bounds} and $L(\hat{A}^{'})=1$ guarantee that $card(Y_{min}^{2})\leq 2$ is sufficient to satisfy Lemma $\ref{le.3cond}$. It can be completed in time at most $n^{2}\times O(n^{2.5})=O(n^{4.5})$.
\end{remark}

\begin{remark}
    Following Corollary \ref{cor2}, Lemma \ref{lem.selfloop}, and Theorem \ref{th.selfloop}, when sensors can measure inputs (allowing $D\ne 0$), the above-mentioned special case can also be solved in polynomial time using Algorithm 1 by omitting the step 1 and replacing $\hat{A}^{'}$ with $A^{'}$.
\end{remark}

	\subsection{Two-stage Heuristic Algorithm for General Case of $\mathcal{P}$}
	A two-stage heuristic algorithm, illustrated in Algorithm \ref{alg}, is proposed based on Lemma \ref{le.cond}, which divides the conditions into two parts. Since achieving \textit{2)} of Lemma \ref{le.cond} is proven to be NP-hard, the second stage of the algorithm can only be suboptimal within polynomial time. Therefore, this two-stage algorithm is an optimal-suboptimal algorithm that can be completed in polynomial time. Define $f(\mathcal{D}(\mathcal{B}^{'}(A,B,C)))$ as the number of subgraphs $\mathcal{B}^{'}_i(i=1,\cdots,k)$ that contains \textit{s-edges} in $\mathcal{D}(\mathcal{B}^{'}(A,B,C))$.
	
	\begin{algorithm}[b]
		\caption{: A two-stage algorithm for $\mathcal{P}$}\label{alg}
		\begin{algorithmic}[1]
			\REQUIRE The structured system $(A,B)$
			\ENSURE A sensor placement $C$
			\STATE Initialize $\mathcal{X}=\{x_1\cdots,x_n\}$ as the set of states, $\mathcal{Y}=\emptyset$ as the set of states which are measured by sensors, $C=0$.
			\STATE Find a maximum matching $\mathcal{M}$ matching all inputs in $\mathcal{B}(A,B)$.
			\FOR{$i=1$ to $n$} 
			\IF{$x_i$ is left-unmatched with $\mathcal{M}$}
			\STATE Update $\mathcal{Y}=\mathcal{Y}\cup\{x_i\}$, $\mathcal{X}=\mathcal{X} \backslash \{x_i\}$, $C=[C;I^{i}]$.
			\ENDIF
			\ENDFOR
			\STATE Denote $\mathcal{J}$ as the set of the subscripts of $x_i\in \mathcal{X}$.
			\WHILE{$f(\mathcal{D}(\mathcal{B}^{'}(A,B,C)))>0$}
			\FOR{$j\in \mathcal{J}$}
			\STATE $\alpha_j=f(\mathcal{D}(\mathcal{B}^{'}(A,B,C)))-f(\mathcal{D}(\mathcal{B}^{'}(A,B,[C;I^{j}])))$.
			\ENDFOR
			\STATE $j_{max}\leftarrow$arg max$_{j\in \mathcal{J}}$ $\alpha_j$.
			\STATE Update $\mathcal{Y}=\mathcal{Y}\cup\{x_{j_{max}}\}$, $\mathcal{X}=\mathcal{X} \backslash \{x_{j_{max}}\}$, $\mathcal{J}=\mathcal{J}\backslash \{j_{max}\}$, $C=[C;I_{j_{max}}]$.
			\ENDWHILE
			\RETURN Sensor placement $\mathcal{Y}$ and output matrix $C$.
		\end{algorithmic}
	\end{algorithm}
	
	\subsubsection{Optimal Algorithm for the First Stage}
	In the first stage, sensors $C_1$ are placed for $(A,B)$ to achieve \textit{1)} of Lemma \ref{le.cond}. Specially, achieving \textit{1)} is equivalent to ensuring that the maximum matching of $\mathcal{B}(A,B,C_1)$ is left-perfect. This task can be easily accomplished using a maximum matching algorithm in polynomial time \cite{murota1987system}. To be specific, we just need to find a maximum matching $\mathcal{M}$ of $\mathcal{B}(A,B)$, where all input vertices are matched with $\mathcal{M}$. Then, sensors are added to the unmatched state vertices associated with $\mathcal{M}$. The main computational complexity at this stage lies in finding the maximum matching, which can be solved in time $O(N^{0.5}M)=O(n^{2.5})$ time, where $N$ is the number of state and input vertices, and $M$ is the number of edges incident to them \cite{murota2010matrices}. This stage corresponds step $1$ to $7$ in Algorithm~ \ref{alg}.
	
	\subsubsection{Suboptimal Algorithm for the Second Stage}
	Building upon the first stage, the second stage aims to add additional sensors $C_2$ for $(A,B,C_1)$ to satisfy \textit{2)}. As shown in Theorem \ref{th.2)} and the proof of Theorem \ref{th.NP}, this stage can be cast into a set cover problem. Consequently, a greedy heuristic is employed to find suboptimal solutions for this stage in polynomial time. Indeed, the greedy algorithm has been proven to achieve the best approximation bounds in polynomial time for the set cover problem \cite{murota1987system}. The complexity of DM-decomposition is $O(MN^{0.5})=O(n^{2.5})$ since its main components involve finding the maximum matching and SCC decomposition \cite{murota2010matrices}. Thus, the second stage can be completed in $O(n^{2.5}\times n{\rm log}(n))=O(n^{3.5}{\rm log}(n))$ time, where $(n{\rm log}(n))$ accounts for the greedy algorithm. This stage corresponds to step $8$ to $15$ in Algorithm~\ref{alg}.
	
	To sum up, we can resort to the maximum matching algorithm in the first stage, while the second stage utilizes a greedy heuristic to obtain optimal or suboptimal solutions within each stage. The overall complexity of Algorithm 2 is $O(n^{3.5}{\rm log}(n))$. Moreover, the second stage relies heavily on the sensor placement of the first stage. Therefore, while this two-stage algorithm provides efficient solutions, it does not guarantee optimal solutions.
	
\section{Illustrative example}\label{ex-sec}

\subsection{An Illustrative Example of the Polynomially Solvable Case}\label{Exam.1}

\begin{figure}[t]
		\centering
		\includegraphics[width=9cm]{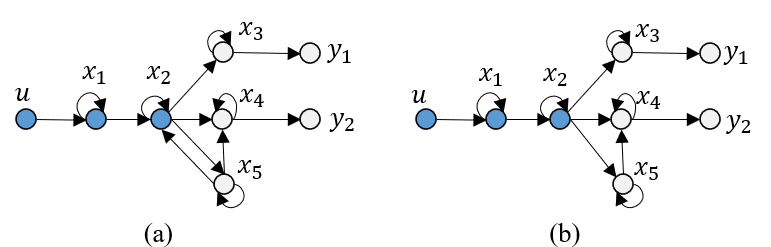}
		\caption{The network topology of $(A^{'},e_{i},C_{min})$ and $(\hat{A}^{'},e_{i},C_{min})$ in Subsection \ref{Exam.1}. Blue vertices represent $\mathcal{V}_{ess}(\{u\},Y_{min})$. }
		\label{fig.ex1}
	\end{figure}

Consider a network $(A^{'},e_{i})$ depicted in Fig. \ref{fig.ex1}. Based on Algorithm \ref{alg.self}, we obtain $Y_{min}=\{y_{1},y_{2}\}$ with output matrix being $C_{min}$. $(A^{'},e_i,C_{min})$ is GSIO, which can be verified by Lemma \ref{le.3cond}. As illustrated in Fig. \ref{fig.ex1}, $\theta(\{u\}\cup X,X\cup Y_{min})=6$. Moreover, input and state vertices are $Y$-reached
. Additionally, $\mathcal{V}_{ess}(\{u\},Y_{min})=\Delta_0=\{x_{1},x_2\}$. Therefore, $(A^{'},e_{i},C_{min})$ is GSIO.

\subsection{An Illustrative Example of General Case}\label{Exam.2}

	\begin{figure}[t]
		\centering
		\includegraphics[width=9cm]{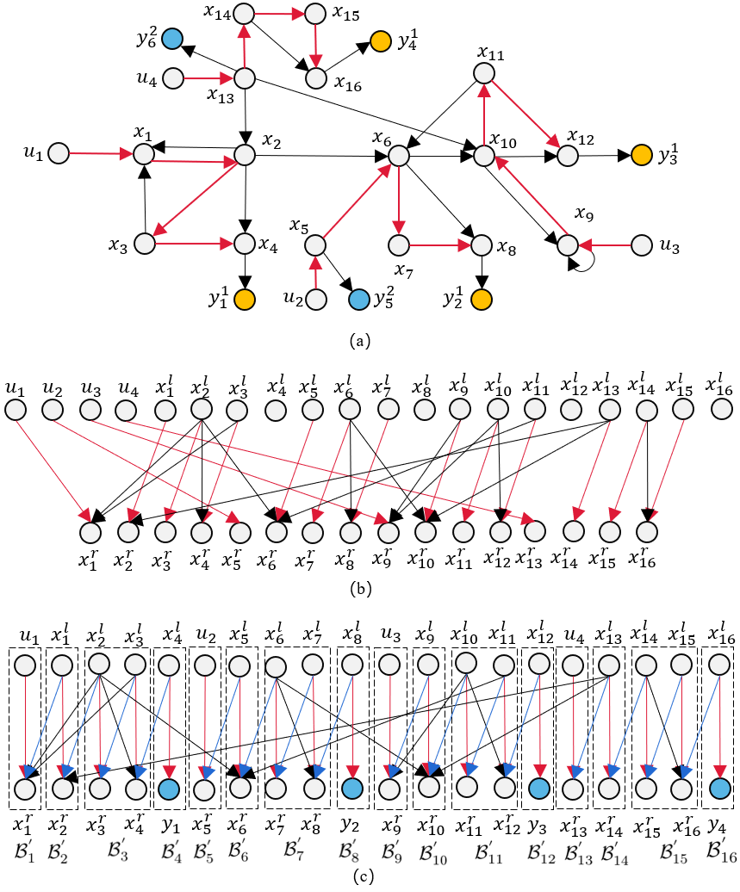}
		\caption{(a) The network topology of $(A,B)$ in Subsection \ref{Exam.2}. (b) The bipartite digraph of $(A,B)$. (c) The DM-decomposition of $(A,B,C_1)$ with $\mathcal{B}_{3}^{'}$, $\mathcal{B}_{7}^{'}$, $\mathcal{B}_{11}^{'}$, $\mathcal{B}_{15}^{'}$ containing \textit{s-edges}. Red lines represent maximum matching. Blue lines represent \textit{s-edges}. Yellow and blue vertices represent sensors placed at the first stage and the second stage, respectively. }
		\label{fig.ex}
	\end{figure}
	
	Consider a network $(A,B)$ with 4 inputs and 16 states, depicted in Fig. \ref{fig.ex}. In the first stage of sensor placement, 4 sensors $C_1$ are placed to measure $\{x_4, x_8, x_{12}, x_{16}\}$, respectively. Importantly, $C_1$ is the minimum sensor placement required to fulfill \textit{1)}, as $card(\mathcal{V}^l)-card(\mathcal{V}^r)=4$ of $\mathcal{B}(A,B)$ in Fig. \ref{fig.ex}. Following the DM-decomposition of $\mathcal{B}^{'}(A,B,C_1)$, 4 subgraphs $\mathcal{B}^{'}_i$ containing \textit{s-edges} are identified. According to Theorem \ref{th.2)} and \ref{th.NP}, the second stage transforms into an extended set cover problem, which can be described as follows. The sets involved are denoted as $S=\{\mathcal{B}^{'}_3,\mathcal{B}_7^{'},\mathcal{B}_{11}^{'},\mathcal{B}_{15}^{'}\}$, with subsets $S_1=\{\mathcal{B}^{'}_3,\mathcal{B}^{'}_{15}\}$ corresponding to $\mathcal{B}^{'}_{2}$, $S_2=\{\mathcal{B}^{'}_3,\mathcal{B}_7^{'},\mathcal{B}^{'}_{11}\}$ corresponding to $\mathcal{B}^{'}_{6}$, $S_3=\{\mathcal{B}^{'}_7, \mathcal{B}^{'}_{11},\mathcal{B}^{'}_{15}\}$ corresponding to $\mathcal{B}^{'}_{10}$, $S_4=\{\mathcal{B}^{'}_{15}\}$ corresponding to $\mathcal{B}^{'}_{14}$, $S_5=\{\mathcal{B}^{'}_3\}$ corresponding to $\mathcal{B}^{'}_{3}$, $S_6=\{\mathcal{B}^{'}_7\}$ corresponding to $\mathcal{B}^{'}_{7}$, $S_7=\{\mathcal{B}^{'}_{11}\}$ corresponding to $\mathcal{B}^{'}_{11}$ and $S_8=\{\mathcal{B}^{'}_{15}\}$ corresponding to $\mathcal{B}^{'}_{15}$. While $S_4=S_8$, it is crucial to note that the corresponding sensor placements are different. Utilizing the greedy heuristic, the process involves initially selecting $S_2$ (resp. $S_3$), followed by the choice of $S_1\text{ or }S_3\text{ or }S_4\text{ or }S_8$ (resp. $S_1\text{ or }S_2\text{ or }S_5$). Consequently, the two-stage heuristic algorithm yields a minimum sensor placement of 6 sensors for $(A,B)$, providing an optimal-suboptimal solution to $\mathcal{P}$.
	\section{Conclusion}\label{con-sec}
   	This paper investigates the problem of determining the minimum number of dedicated sensors required to achieve GSIO for structured systems. We revisit and refine existing sufficient and necessary conditions for GSIO with respect to sensor additions. Based on the new conditions, we demonstrate the NP-hardness of this problem. Moreover, we propose an upper bound and a lower one when inputs are dedicated, relating the minimal GSIO problem to the extensively studied minimal structural observability problem. Additionally, we present a special case for which the exact optimal value can be determined in polynomial time. Finally, we propose a two-stage algorithm in the general case. Each stage of the algorithm is designed to be either optimal or suboptimal and can be completed in polynomial time.
    Given the recent advances on the observability of partial states \cite{fernando2010functional,boukhobza2014partial, zhang2023functional}, in the future we plan to investigate the sensor placement problem for achieving partial observability of a given set of states in the presence of unknown inputs \cite{boukhobza2009structural, boukhobza2010partial}, which is more attractive and challenging.




	\bibliographystyle{elsarticle-num}
	{\small
		\bibliography{ISO-journal-2}
	}
\end{document}